\documentclass[11pt]{article}

\usepackage[margin=2.4cm]{geometry}
\usepackage{amsmath,amssymb,amsthm}
\renewcommand{\boxed}[1]{{\displaystyle #1}}
\usepackage{graphicx}
\usepackage[numbers,sort&compress]{natbib}
\usepackage{booktabs}
\usepackage{longtable}
\usepackage{caption}
\usepackage{xcolor}
\definecolor{linkc}{RGB}{30,70,130}
\usepackage[colorlinks=true,allcolors=linkc,citecolor=red]{hyperref}
\usepackage{microtype}
\hypersetup{
  pdftitle={Selection rules for the harmonic spectroscopy of animal decisions},
  pdfauthor={Mohammad Salahshour},
  pdfsubject={Harmonic spectroscopy of animal decisions}
}

\newtheorem{proposition}{Proposition}
\newcommand{\figpanel}[2]{Fig.~\ref{#1}#2}
\graphicspath{{./}}

\title{\bfseries Selection rules for the harmonic spectroscopy of animal decisions}

\author{Mohammad Salahshour$^{1,2,3,\ast}$\\[4pt]
\small $^{1}$Department of Collective Behavior, Max Planck Institute of Animal
Behavior, Konstanz, Germany\\
\small $^{2}$Centre for the Advanced Study of Collective Behaviour, University of
Konstanz, Konstanz, Germany\\
\small $^{3}$Department of Biology, University of Konstanz, Konstanz, Germany\\[4pt]
\small $^{\ast}$Correspondence: \texttt{salahshour.mohammad@gmail.com}}
\date{}

\begin{document}
\maketitle

\begin{abstract}
Spectroscopy reads structure by sending a structured probe into a system and
measuring what comes back.  Here we apply this principle to animal
decision-making.
Harmonic Theory casts choice as motion on an angular landscape over heading,
whose Fourier components form an animal's decision spectrum.  We show that the
arrangement of cues enters that landscape as a structure factor, so it factors
like a diffraction amplitude, and symmetry imposes selection rules: a $p$-fold
cue array annihilates every harmonic that is not a multiple of $p$, and
sweeping two cues apart draws a notch through each harmonic.  Cue geometry
becomes a tunable probe: the experimenter sets which harmonics an experiment
can see.  We read these spectra from public recordings of flies, schooling
fish, larval zebrafish, mice and a swimming alga, and from the fly's neural
compass, where symmetric scenes suppress the forbidden harmonics and breaking
that symmetry restores them.  Measuring the compass form factor favours the
theory's smooth bump over its square one, and raising cue contrast narrows
that bump in every fly.  A width fitted to one harmonic then predicts the
next.  Cue geometry is a spectrometer for decisions, and symmetry sets what it
can see.
\end{abstract}

\section{Introduction}

Animal navigation is a mapping from the angular arrangement of the world onto
a direction of travel.  A moving animal is surrounded by neighbours,
landmarks, food patches and sources of light, and at every moment it must turn
their bearings into a single heading
\citep{salahshour2025allocentric,seelig2015neural,karpenko2020light,sridhar2021geometry}.  Many experiments probe this
mapping by presenting a controlled arrangement of cues and measuring the
animal's turning response.  This approach has been used with flies steering among virtual
targets \citep{sridhar2021geometry}, with locusts marching in a swarm
\citep{sayin2025behavioral}, with larval zebrafish responding to conspecifics
and to light \citep{harpaz2021precise,harpaz2024experience,karpenko2020light},
with schooling fish under manipulated illumination and virtual neighbours
\citep{xue2023light,mckee2020sensory,li2025biopd}, and with a unicellular alga
steering between two beams \citep{raikwar2025phototactic}.  One regularity
recurs.  The geometry of the options organizes the response, and beyond a
critical angular separation an animal stops compromising between two options
and commits to one --- in flies, in locusts and in larval zebrafish alike
\citep{sridhar2021geometry,sayin2025behavioral}.  The transition appears
across taxa and across scales of organization, which suggests that something
general is being measured.  What that something is, in the sense of a quantity
with units that one could estimate from data and carry from one experiment to
another, has been much less clear.

Harmonic Theory supplies a candidate \citep{salahshourcouzin2026harmonic}.  If a
directional variable is represented on a ring of neurons, if sensory input adds
linearly, and if the decision state is a localized activity bump, then the
influence of a cue on heading is a periodic landscape whose Fourier
coefficients factor into an interaction strength, a sensory-width filter, and a
neural form factor --- the Fourier spectrum of the localized activity bump
on the ring.  Each Fourier coefficient indexes a \emph{harmonic}: the
first harmonic is a simple pull toward a direction, the second prefers an
axis, and higher harmonics encode progressively finer angular structure.
Ring-attractor dynamics with these qualitative features are measured in the
\emph{Drosophila} central complex, where a localized bump tracks heading, is
anchored by visual cues, and combines several sensory cues into one heading
estimate \citep{seelig2015neural,kim2017ring,basnak2025multimodal}; how the
width of that bump responds to cue contrast is measured here.  The theory
therefore proposes a specific bridge from a neural representation to a
behavioral force law, with the harmonic coefficients $K_n$ providing the link.

Measuring that spectrum remains a challenge despite the available data.  Methods that
infer turning rules from trajectories recover rich response maps, often anisotropic
and non-central
\citep{katz2011inferring,herbertread2011inferring,calovi2018disentangling},
but those maps are rarely decomposed into angular harmonics, so the
higher-harmonic content of a response goes unmeasured.  When such a rule is
summarised compactly, the summary is usually an attraction toward the average,
or resultant, direction of the options --- a first-harmonic response
\citep{harpaz2021precise,li2025biopd}.  One harmonic is a
severe restriction.  For instance, in the compromise-to-selection transition, a
landscape containing only $n=1$ has a stable bisector at every separation short
of antipodal --- at a fixed symmetric cue geometry with equal cue gains --- so
it cannot produce a local instability there at all
\citep{sridhar2021geometry}.  Such transitions are usually captured by positing
a separate winner-take-all or bistable mechanism alongside the geometric
response \citep{sridhar2021geometry,raikwar2025phototactic},
rather than by measuring the harmonic content that would produce them.  The
transition alone does not settle this: as we show here, pooling frames whose
cue gains depend on distance can split the observed distribution as well, and
a first-harmonic model refitted to the data still predicts the two-target
transition too late, producing none at all in the three-target cohort.

This paper argues that the discrepancy --- a theory that predicts a spectrum,
a literature that reports one harmonic --- is a measurement problem with a
solvable structure, and it provides four components: a design theory that identifies
which harmonics an experiment can see, a second estimator that sees them
better, a behavioural demonstration in existing public recordings, and a
neural test of the same selection rules in the
fly's compass.

The instrument comes from writing the cue configuration as a complex
\emph{structure factor} $F_n=\sum_j w_j e^{-\mathrm i n\Theta_j}$ --- one
complex number per harmonic, calculated from the cue bearings and weights at
each observation.  The landscape then factors into an intrinsic animal spectrum times this
known geometric quantity, the same factorization that splits a diffraction
amplitude into what scatters and how it is arranged
\citep{alsnielsen2011elements}.  Harmonic Theory does not resemble a spectral
theory; it is one.  Consequently, methods from spectroscopy can be applied
directly to the analysis of animal decisions.  An extended cue and the neural activity bump
each contribute a \emph{form factor}, the filter set by the shape of one
scatterer as against the arrangement of many.  Symmetry imposes
\emph{selection rules}: a cue array invariant under rotation by $2\pi/p$
annihilates every harmonic that is not a multiple of $p$, as symmetry forbids
transitions in atomic and molecular spectra \citep{bernath2016spectra}.  Three
targets at $120^\circ$ delete the first and second cue-locked harmonics
identically, so any orientation structure such an array can produce begins at
the third.  Sweeping a pair's separation sweeps a comb of notches through the
harmonics, switching each off at an angle known in advance.  A coefficient
that a symmetry zero renders unmeasurable is then not a failed regression but
a \emph{systematic absence}, the object that carries the symmetry information
in crystallography \citep{ita2016spacegroup}.  And a closed-form detectability
law, Eq.~\eqref{eq:detectability}, says how much detectable signal survives at
each order: given a broadly tuned sensory front end, it predicts the
first-harmonic ceiling of the literature, and it says how much observation
time or geometric leverage would be needed to break it.  Cue geometry
becomes a spectrometer and the animal is the sample, with symmetry fixing in
advance which lines the instrument can pass.  We call the method harmonic
spectroscopy, and the field it opens --- measuring an animal's response
spectrum by choosing the geometry of the cues it is given --- animal
spectroscopy.

We apply this approach to existing recordings.  Flies steering
between two swept virtual targets reverse their second harmonic near
$91^\circ$, where that design's structure factor puts its notch; a three-target
array moves the predicted notch to about $120^\circ$, and the measured reversal
moves with it.  The sharpest test is neural, because symmetry can be printed
directly onto a visual scene: panoramas with two- and four-fold symmetry
suppress the harmonics their symmetry forbids, symmetry-broken twins restore
them in the same flies, and every prespecified discriminator shifts in the
predicted direction in every animal.  The selection rules that organize the
behaviour are read out in the neural representation the behaviour is supposed
to inherit from --- and the same recordings measure the neural filter the theory had
only assumed \citep{salahshourcouzin2026harmonic}, finding it smooth. Not everything
transfers: fitted higher harmonics do
not carry between the two cue designs, the two channels agree in shape but not
in absolute scale, and separability is not what we offer as evidence, since
separable kernels already fit fish interaction data
\citep{calovi2018disentangling}.  What this paper contributes is not another
rule for how animals choose a direction, but a way to measure the one they
already use --- and a law that says in advance which harmonics an experiment
can see.

\section{Results}
\label{sec:results}

\subsection{Cue geometry as a diffraction grating}
\label{sec:diffraction}

\subsubsection{The measurement equation}

Let $\Phi_t\in\mathbb S^1$ denote an animal's heading and let a set of cues
$j=1,\dots,J$ have allocentric bearings $\Theta_{jt}$, distances $d_{jt}$ and
weights $w_{jt}$.  Harmonic Theory writes the instantaneous decision landscape
over heading as a sum of localized periodic wells
\citep{salahshourcouzin2026harmonic},
\begin{equation}
  H(\phi,t)=-\sum_{j}w_{jt}\sum_{n\ge1}K_n(d_{jt})
    \cos\!\big[n(\phi-\Theta_{jt})\big],
  \label{eq:landscape}
\end{equation}
with the factorized spectrum
\begin{equation}
  K_n(d)=J_{\mathrm{int}}(d)\,c_n(\sigma)\,M_n(W),\qquad
  c_n(\sigma)=\frac{1}{\pi}e^{-n^2\sigma^2/2},\qquad
  M_n(W)=\frac{4}{n\pi}\sin\!\Big(\frac{nW}{2}\Big),
  \label{eq:factorization}
\end{equation}
where $J_{\mathrm{int}}(d)$ sets the overall strength of the response to a cue
at distance $d$ --- a common gain at every harmonic
--- while $\sigma$ is the width of the sensory kernel on the ring and $W$ the
support width of the localized neural decision state, the activity bump.

The cue contributions can be combined through the
\emph{complex cue structure factor}
\begin{equation}
  \boxed{\;F_n(t)=\sum_{j}w_{jt}\,e^{-\mathrm i n\Theta_{jt}}\;}
  \label{eq:structure-factor}
\end{equation}
If the harmonic kernel is common to the cues present at time $t$ --- exactly
true when they are equidistant, and controllable otherwise by stratifying on
distance --- then Eq.~\eqref{eq:landscape} collapses to
\begin{equation}
  H(\phi,t)=-\sum_{n\ge1}K_n\,\mathrm{Re}\!\left[F_n(t)\,e^{\mathrm i n\phi}\right].
  \label{eq:landscape-sf}
\end{equation}
The experiment enters only through $F_n$, a known complex number fixed by the
cue arrangement; the animal enters only through the real coefficients $K_n$
(\figpanel{fig:selection}{a,b}).
Eq.~\eqref{eq:landscape-sf} is the measurement equation of harmonic
spectroscopy, and it has the algebraic form of a diffraction amplitude: an
intrinsic spectrum multiplied by a structure factor that encodes the
arrangement of scatterers \citep{alsnielsen2011elements}.

When cues are extended rather than point-like, the analogy is exact.  A cue
subtending a \emph{full} angular width $\Theta^{\mathrm{sub}}$ contributes an
additional \emph{form factor} --- the filter set by the shape of one cue, as
distinct from the arrangement ---
$S_n(\Theta^{\mathrm{sub}})=\tfrac{4}{n\pi}
\sin(n\Theta^{\mathrm{sub}}/2)$, whose first zero is at
$\Theta^{\mathrm{sub}}=360^\circ/n$, so that
$F_n=\sum_j w_j S_n(\Theta^{\mathrm{sub}}_{jt})e^{-\mathrm i n\Theta_{jt}}$.
Here $w_j$ is a contrast-density weight, so total cue drive grows with angular
extent.  If $w_j$ instead denotes fixed total cue weight, the normalized form
factor is $\sin(n\Theta^{\mathrm{sub}}/2)/(n\Theta^{\mathrm{sub}}/2)$; the zeros
and within-condition harmonic ratios are unchanged.  Arrangement and
individual shape separate exactly as in scattering theory
\citep{alsnielsen2011elements}.  Supplementary Note~S1 gives the integral and
explains why a cue size shared by all cues is not identifiable without further
constraints or a size manipulation.

It is tempting to conclude that $S_n$ and the neural filter $M_n(W)$ share a
functional form and that cue size and bump width are therefore degenerate.
That holds only for a square bump, a shape the neural data do not support.
The two filters differ over the harmonics resolved here: $S_n$
describes a sharp-edged object and so has \emph{zeros}, changing sign at
$\Theta^{\mathrm{sub}}=360^\circ/n$, whereas the measured compass profile
decays monotonically and changes sign at no resolved harmonic
($M_n/M_1=1,\,0.41,\,0.11$ at high contrast, in the bump's own frame;
Sec.~\ref{sec:contrast}).  The smooth kernel families supported by these data
have no zeros over those resolved orders.

Zeros can therefore break the degeneracy, but only with an additional
assumption or manipulation.  If the downstream sensory and neural filters are
non-zero and sign-preserving at harmonic $n$, a sign change there indexes the
angular size of a sharp-edged object,
$\Theta^{\mathrm{sub}}=360^\circ/n$.  Without that condition, or a controlled
cue-size manipulation, only the product of the cue, sensory and neural filters
is identifiable (Sec.~\ref{sec:general-kernel}; S1).

\subsubsection{A general kernel, and what is identifiable}
\label{sec:general-kernel}

Eq.~\eqref{eq:factorization} writes $M_n$ for a square activity bump; Harmonic
Theory gives the Gaussian case as well. Nothing in the formalism requires either, so
we state the general version and keep both as special cases.

Every stage between a cue bearing and a turning drive acts on the ring by
convolution: the object's angular extent, the sensory kernel, the
neural bump \citep{salahshourcouzin2026harmonic}.  On the circle, convolution
multiplies Fourier coefficients, so
\begin{equation}
  K_n(d)=J_{\mathrm{int}}(d)\;\hat s_n\,\hat e_n\,\hat b_n ,
  \label{eq:general-factorization}
\end{equation}
in which $\hat s_n$ belongs to the cue's own angular profile --- it is the form
factor $S_n(\Theta^{\mathrm{sub}})$ above, now written for an arbitrary shape
rather than a sharp-edged arc --- $\hat e_n$ to the sensory kernel,
and $\hat b_n$ to the neural bump.  Up to the normalization each carries, they
are characteristic functions $\mathbb E[e^{-\mathrm i nX}]$ --- the Fourier
coefficients of the random angular offset $X$ each stage adds --- so the
spectrum is a product of characteristic functions. Harmonic Theory's factorization
is the case $\hat s_n\hat e_n=c_n(\sigma)$ with $\hat b_n$ square or Gaussian.

For a Gaussian bump the coefficient follows in one step, and it is the
case the neural data are fitted with.  A bump that is normal with width
$\sigma_{\mathrm{bump}}$ before wrapping has the circular density
$b(\theta)=\sum_k(2\pi\sigma_{\mathrm{bump}}^2)^{-1/2}
\exp[-(\theta+2\pi k)^2/2\sigma_{\mathrm{bump}}^2]$.
Because $e^{-\mathrm i n\theta}$ has period $2\pi$, the sum over windings unfolds
the integral over one period onto the whole line,
\[
  \hat b_n=\int_{-\pi}^{\pi}b(\theta)e^{-\mathrm i n\theta}\,\mathrm d\theta
  =\int_{-\infty}^{\infty}
    \frac{e^{-x^2/2\sigma_{\mathrm{bump}}^2}}{\sqrt{2\pi}\,\sigma_{\mathrm{bump}}}
    e^{-\mathrm i nx}\,\mathrm dx
  =e^{-n^2\sigma_{\mathrm{bump}}^2/2}.
\]
Wrapping therefore leaves the coefficients unchanged at integer frequencies:
they are the characteristic function of the unwrapped normal, evaluated at $n$.  Normalising,
$\hat b_n/\hat b_1=e^{-(n^2-1)\sigma_{\mathrm{bump}}^2/2}$.  Three consequences
follow.

\paragraph{Identifiability.}
Only the product in Eq.~\eqref{eq:general-factorization} is observable.  If the
smooth stages are wrapped normal --- Gaussian, wrapped onto the circle --- with
widths $\sigma$ and $\sigma_{\mathrm{bump}}$, their product is wrapped normal with width
$\sqrt{\sigma^2+\sigma^2_{\mathrm{bump}}}$, so a behavioural spectrum
identifies that combination and never either term.  The degeneracy is exact,
not a limitation of estimation: no quantity of behavioural data resolves it.

\paragraph{Composition.}
Write the wrapped stable family as the harmonic ratio
$\hat b_n/\hat b_1=e^{-(|n|^\alpha-1)\gamma^\alpha/2}$, with width $\gamma$ and
exponent $\alpha$ --- normalised to the first harmonic, so it is not itself a
characteristic function.  Convolution preserves $\alpha$ and adds the
$\alpha$-th powers of the widths; ``variances add'' is the case $\alpha=2$,
the wrapped normal derived above.
If the cue, sensory and neural filters have different decay exponents $\alpha$,
their product no longer belongs to the same wrapped-stable family.  A change in
the fitted $\alpha$ when one filter is manipulated can therefore reflect
filters with different decay profiles, but can also arise from a change in
an individual filter's shape.
The neural recordings resolve enough harmonics to estimate $\alpha$, as done in
Sec.~\ref{sec:contrast} and S7.

\paragraph{The transition depends on shape, not only width.}
To see why kernel shape matters, consider two identical cues at bearings
$-\Delta/2$ and $+\Delta/2$ relative to their bisector
\citep{salahshourcouzin2026harmonic}.  Their separation is $\Delta$, and their
structure factor is
\begin{equation}
  F_n(\Delta)=2\cos\!\Big(\frac{n\Delta}{2}\Big).
  \label{eq:pair-sf}
\end{equation}
Their midpoint, $\phi=0$, is the bisector --- the direction exactly between
them and therefore the compromise heading in the decision landscape of
Eq.~\eqref{eq:landscape-sf}.  The bisector is a stable valley when its curvature
$C(\Delta)=H_\Delta''(0)$ is positive and becomes an unstable ridge when that
curvature is negative; the critical separation $\Delta_c$ is where
$C(\Delta_c)=0$.  The $n$th term enters the landscape with coefficient $K_n$,
whereas differentiating it twice weights its contribution to the curvature by
$n^2$.  Higher harmonics therefore strongly influence the transition.

The persistence of those harmonics is determined by the complete decision
spectrum $K_n\propto\hat s_n\hat e_n\hat b_n$, which combines cue shape, the
sensory kernel and the neural bump (Eq.~\eqref{eq:general-factorization}).
Kernels of equal width can therefore produce different transitions if their
harmonics decay differently.  Approximating $K_n/K_1$ by the wrapped-stable
family, the wrapped Cauchy ($\alpha=1$) retains higher harmonics much more
strongly than the wrapped normal ($\alpha=2$).  The effect is large: at
$\gamma=0.8$~rad, $\Delta_c$ shifts from $26^\circ$ to $92^\circ$ (S1).
Stable laws are defined for $0<\alpha\le2$ \citep{nolan2020stable}.  Extending
the displayed coefficient formula beyond that family is not a stable-law model.
If the
sensory kernel is instead held Gaussian at $\sigma=0.8$~rad and the neural
filter is a square bump, increasing its width $W$ moves $\Delta_c$ only upward,
from $92^\circ$ towards $180^\circ$, and eliminates the transition for
$W\ge180^\circ$.  A square neural bump therefore cannot produce a critical
separation below the wrapped-normal value (S1).

\subsubsection{Selection rules}

\begin{proposition}[Symmetry selection rule]
\label{prop:selection}
If the weighted cue configuration is invariant under rotation by $2\pi/p$ for
an integer $p\ge2$, then $F_n=0$ unless $p\mid n$.
\end{proposition}

\begin{proof}
Rotating every bearing by $2\pi/p$ maps $F_n\mapsto e^{-2\pi\mathrm i n/p}F_n$
while leaving the configuration, and hence $F_n$, unchanged.  Therefore
$F_n\big(1-e^{-2\pi\mathrm i n/p}\big)=0$, and the bracket vanishes only when
$p\mid n$.
\end{proof}

\begin{figure}[p]
\centering
\includegraphics[width=\textwidth]{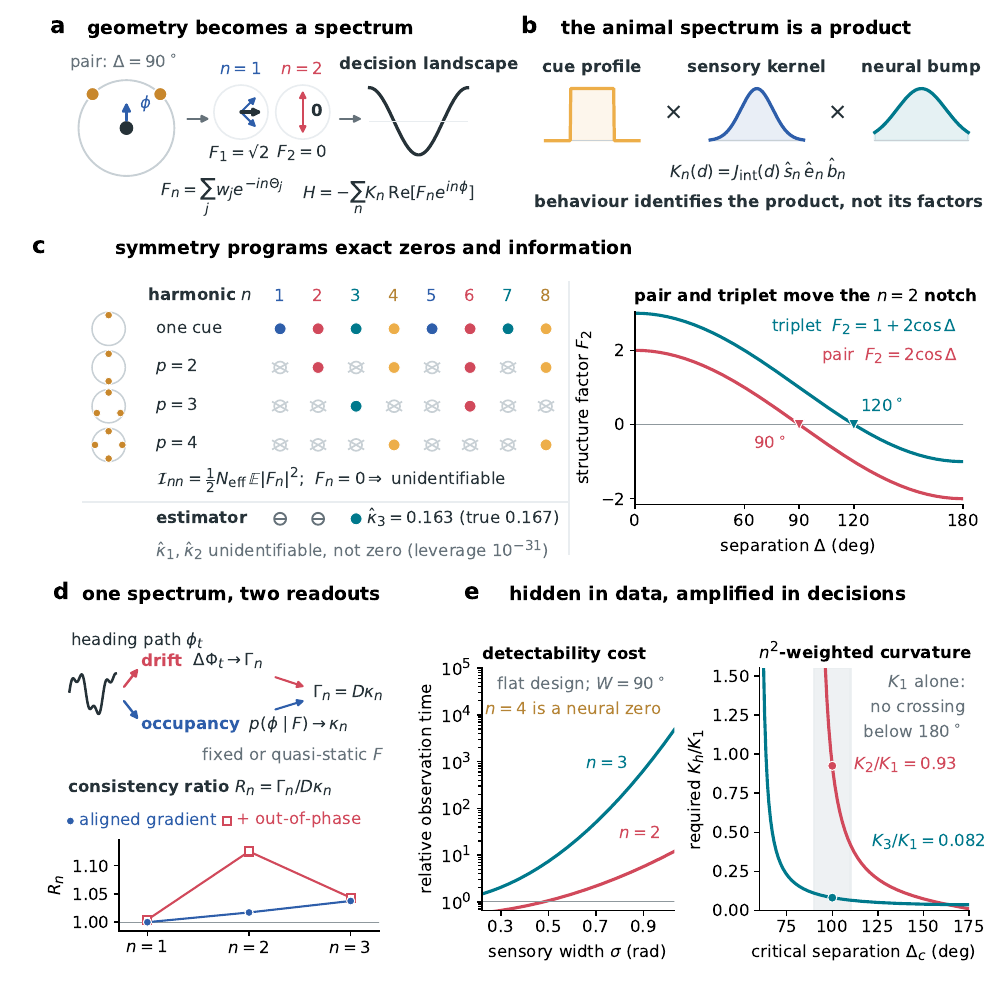}
\caption{
\textbf{Theory map of harmonic spectroscopy.}
Closed form and a validation simulation; no animal data enter.
\textbf{a}, Cue geometry fixes the complex structure factor $F_n$ and hence
the decision landscape: for the $90^\circ$ pair the first-harmonic phasors add
while the second cancel, leaving only $n=1$ in the illustrated two-harmonic
landscape.  \textbf{b}, Cue
shape, sensory filtering and the neural bump multiply in harmonic space;
behaviour identifies their product, not its factors.  \textbf{c}, A regular
$p$-fold array imposes exact symmetry zeros: filled circles allowed harmonics,
crossed circles forbidden ones, though an allowed multiple can cancel in a
less regular array.  Fisher information follows $|F_n|^2$, so a forbidden
coefficient is unidentifiable rather than zero --- what the estimator row
below the grid returns.  The neural symmetry test is shown in
\figpanel{fig:kim}{b--d}.  Right, pair and triplet sweeps put the
second-harmonic zero at $90^\circ$ and $120^\circ$, tested in
\figpanel{fig:bifurcation}{a}.  \textbf{d}, One
trajectory gives a drift and an occupancy readout; under the cue-aligned
gradient model with constant diffusion and equilibrated geometry,
$\Gamma_n=D\kappa_n$.  The inset tests agreement and the response to an
out-of-phase term at $n=2$.  \textbf{e},
Higher harmonics are harder to detect yet carry $n^2$ weight in curvature:
left, detectability for a flat design and square bump, $W=90^\circ$; right,
the ratios a critical separation needs.
The two measured readouts (\textbf{c}, \textbf{d}) come from the validation
record of Note~S1: $4000$ trajectories, $4000$ steps,
$\mathrm{d}t=0.01$~s.}
\label{fig:selection}
\end{figure}

A $p$-fold symmetric cue array is thus an exact harmonic comb filter: it
annihilates every harmonic that is not a multiple of $p$ ---
including those above $p$ --- and passes $n=p,2p,\dots$
(\figpanel{fig:selection}{c}).  For one orbit of $p$
equally spaced cues, the passed amplitude is $|F_{p}|=\sum_j w_j$
(Eq.~\eqref{eq:structure-factor}); a configuration built from several such
orbits at different phases is still annihilated at non-multiples of $p$, but its
passed harmonics can partially cancel, leaving $|F_p|$ smaller.  Three identical cues at $120^\circ$
delete the first and second harmonics identically, so \emph{any} orientation
structure that survives in such an array is third-harmonic or higher.  This
inverts the usual reading of a symmetry zero.  In an unconstrained regression, a
predictor that is identically zero --- a vanishing design column --- is a
failure; here it is the instrument.  This is exactly the
situation of systematic extinctions in crystallography, where the absent
reflections carry the symmetry information
\citep{ita2016spacegroup,alsnielsen2011elements}.

The continuous version is more useful still.  For the two-cue geometry of
Eq.~\eqref{eq:pair-sf}, the structure factor vanishes whenever $n\Delta/2$ is
an odd multiple of $\pi/2$, first at $\Delta=\pi/n$.  Sweeping $\Delta$
therefore sweeps a comb of harmonic notches:
at $\Delta=180^\circ$ the first harmonic is switched off (along with every
other odd harmonic, since an antipodal pair is twofold symmetric), at
$90^\circ$ the second, at $60^\circ$ the third.  For a symmetric triplet with a
cue on the bisector, $F_n(\Delta)=1+2\cos(n\Delta/2)$, whose first zeros lie at
$\Delta=4\pi/(3n)$: the second-harmonic notch moves from $90^\circ$ to
$120^\circ$ --- a \emph{different} comb (\figpanel{fig:selection}{c}).  Two such families measured
over the same separations give linearly independent projections of the same
spectrum, so cross-geometry transfer is a real test rather than a restatement.

\subsubsection{Design information}

Write $X_n(\phi,F)=\mathrm{Re}[F_n e^{\mathrm i n\phi}]$ for the harmonic term
of Eq.~\eqref{eq:landscape-sf}.  In the weak-response limit, the heading
distribution is close to uniform and
\begin{equation}
  \mathbb E_{\phi\sim\mathrm{Unif}}\!\left[X_nX_m\right]
  =\tfrac12\,\mathrm{Re}\!\left[F_n\overline{F_m}\right]\delta_{nm},
  \label{eq:orthogonality}
\end{equation}
because a nonzero average requires the total phase $n\pm m$ to vanish.  The
harmonics are therefore mutually orthogonal by construction.

Orthogonality makes the information supplied by a design explicit.  The
coefficients enter the heading law linearly through $X_n$, making the model an
exponential family with the coefficients as natural parameters and $X_n$ as
sufficient statistics (Sec.~\ref{sec:channels}).  Its Fisher information ---
the expected curvature of the log-likelihood --- is the covariance matrix of
these statistics \citep{lehmann1998point}.  In the weak-response limit,
headings are approximately uniform, the $X_n$ have mean zero, and
Eq.~\eqref{eq:orthogonality} gives a diagonal covariance matrix.  The
information about harmonic $n$, with temporal correlations represented by an
effective sample count, is
\begin{equation}
  \mathcal I_{nn}=\tfrac12\,N_{\mathrm{eff},n}\,\mathbb E\big[|F_n|^2\big],
  \label{eq:design-information}
\end{equation}
where $N_{\mathrm{eff},n}$ is the effective number of observations at harmonic
$n$, smaller than the number of frames because successive frames are correlated
and potentially different across harmonics (S1).  Thus information at one harmonic cannot compensate for its
absence at another; if $F_n=0$, that harmonic is unobservable.  Moreover,
for independent observations this is Fisher information and its inverse is
the Cram\'er--Rao variance bound \citep{lehmann1998point}.  For correlated recordings it is an effective
precision expression, not a universal bound; the full covariance must account
for temporal correlations and changing geometry (S1).
We call $\mathbb E[|F_n|^2]$ the \emph{spectral leverage} of a design.  For a
prescribed cue geometry and weight schedule, it is computable before data
collection; when geometry depends on the animal's motion, it must instead be
evaluated under an anticipated trajectory distribution or from the realised
geometry.  It is the quantity a spectroscopic experiment must be built to
control.  For a single cue the leverage is unity
at every $n$; for a pair whose separation is swept uniformly over a full period it
is $2$ at every $n$ (Eq.~\eqref{eq:pair-sf}).  Both designs are spectrally flat, so any decline
of measured amplitude with $n$ is a property of the animal rather than of the
apparatus.  A restricted sweep is not exactly flat, and the sweep analysed
here is restricted: the two-target \emph{Drosophila} cohort of
\citet{sridhar2021geometry} visits separations from $30$ to $180^\circ$, not a
full period (Methods).  Its leverage is therefore computed from the separations
the animals actually experienced, not set to the idealised value; the same is
done for every design below.

\subsection{Two measurement channels and a consistency test}
\label{sec:channels}

Overdamped angular motion on the landscape~\eqref{eq:landscape-sf} obeys
\citep{salahshourcouzin2026harmonic}
\begin{equation}
  \mathrm d\Phi_t=-\eta\,\partial_\phi H(\Phi_t,t)\,\mathrm dt
  +\sqrt{2D}\,\mathrm dB_t
  =-\sum_{n\ge1}n\,\Gamma_n\,\mathrm{Im}\!\left[F_n e^{\mathrm i n\Phi_t}\right]\mathrm dt
  +\sqrt{2D}\,\mathrm dB_t,
  \label{eq:sde}
\end{equation}
with mobility $\eta$, angular diffusion coefficient $D$, Brownian motion $B_t$
and $\Gamma_n=\eta K_n$.  Two distinct statistics of the same trajectory
identify the same spectrum.

\paragraph{Dissipative (drift) channel.}
Each harmonic contributes one predictor to
Eq.~\eqref{eq:sde}, $-n\,\mathrm{Im}[F_ne^{\mathrm i n\phi}]$ evaluated along
the trajectory; these predictors are the columns of the design matrix.
Regressing the observed angular increments on them recovers $\Gamma_n$
directly.  This is the channel used by many existing
interaction-inference methods
\citep{katz2011inferring,herbertread2011inferring,calovi2018disentangling}.

\paragraph{Fluctuation (occupancy) channel.}
Because Eq.~\eqref{eq:sde} is a gradient flow with constant diffusion, the
conditional stationary law of the heading is Gibbs,
\begin{equation}
  \boxed{\;
  \log p(\phi\mid F)=\sum_{n\ge1}\kappa_n\,
    \mathrm{Re}\!\left[F_ne^{\mathrm i n\phi}\right]-\log Z(F),
  \qquad \kappa_n=\frac{\eta K_n}{D}=\frac{\Gamma_n}{D}.\;}
  \label{eq:occupancy}
\end{equation}
This is a conditional circular exponential family whose natural parameters are
the harmonic coefficients in units of the diffusion, and whose sufficient
statistics are the cue-weighted harmonic moments
$X_n=\operatorname{Re}[F_ne^{\mathrm i n\Phi}]$.  It uses every frame rather
than every increment, and it never differentiates an angle.
Eq.~\eqref{eq:occupancy} is the stationary law at fixed $F$, so applying
it to moving cues assumes the bearings change slowly compared with the heading
relaxation time.  Moving-observer simulations (S4) probe, but do not determine,
the error introduced by the quasi-static approximation at the sweep rates
observed experimentally.

\paragraph{The consistency test.}
Both channels estimate the same physical spectrum, so
\begin{equation}
  R_n \;:=\; \frac{\Gamma_n}{D\,\kappa_n}\;=\;1\qquad\text{for every }n
  \label{eq:fdt}
\end{equation}
whenever the angular dynamics are an overdamped gradient flow of the
instantaneous landscape with the stated diffusion.  Eq.~\eqref{eq:fdt} is a
harmonic-resolved fluctuation--dissipation relation for decisions.  It is a
genuine falsification test with no free parameter.  Unlike a rank-one or
separability restriction, it cannot be satisfied by an arbitrary flexible kernel
fitted to the same data, because the two channels are computed from different
functionals of the trajectory (\figpanel{fig:selection}{d}).

The drift channel can also test whether the response is aligned with the
instantaneous cue field.  Estimating both quadratures of the response --- its sine and cosine
components --- defines a complex spectrum.  If the cue frame rotates at an
approximately constant angular speed, a single response delay produces a
phase shift proportional to $n$, the angular analogue of a group delay; a
departure from linearity is a misalignment that no single latency can produce.  A sine-only
design cannot make this distinction at all; it absorbs the effect into an
attenuated amplitude.  Note~S1 states this phase spectroscopy in full, with the
caution that on a ring a cosine quadrature is not by itself evidence of
non-gradient dynamics.

\subsubsection{Efficiency and robustness}

\begin{proposition}[Channel equivalence]
\label{prop:equivalence}
For the one-dimensional gradient diffusion~\eqref{eq:sde} with a fixed cue
configuration $F$, known constant $D$, and continuous error-free observation
over total time $T$, the occupancy
estimator of $\Gamma_n$ attains, in the weak-response limit, the same
asymptotic variance as the continuous-path maximum-likelihood drift estimator,
\begin{equation}
  \mathrm{Var}(\hat\Gamma_n)=\frac{4D}{T\,n^2\,\mathbb E[|F_n|^2]}.
  \label{eq:variance}
\end{equation}
\end{proposition}

The proof, together with the finite-coupling correction to the relaxation rate
that sets how far simulations at weak but nonzero coupling sit above this
floor, is given in Supplementary Note~S1.

Under the conditions of Proposition~\ref{prop:equivalence}, stationary
occupancy is therefore asymptotically as informative as the entire path: no
asymptotic information is lost by discarding the increments.  What is gained
is robustness.  Write the observed heading as
$\hat\Phi_t=\Phi_t+\varepsilon_t$, with an error of scale $s$ independent from
frame to frame and a sampling interval $\delta t$.  The drift regression then
puts a single draw of that error on both sides of its own equation.  Its response is the
observed increment, $\hat\Phi_{t+\delta t}-\hat\Phi_t$, which carries
$-\varepsilon_t$; its predictor is built from the observed heading at that same
instant $t$, so it carries $+\varepsilon_t$.  Regression assumes the predictor
is measured exactly, or at least that its error is uncorrelated with the
response's.  Here the two errors are one draw of one noise, perfectly correlated
and opposite in sign, and the fitted coefficient is biased by order
$s^2/\delta t$.
That bias \emph{diverges as the sampling is refined}, because the drift is the
increment divided by $\delta t$: shortening the interval shrinks the true
increment in proportion, while $\varepsilon_t$ stays whatever size the camera
makes it.  Increasing the sampling rate therefore yields more samples but greater bias.  That
measurement noise corrupts drift and diffusion reconstruction in this way is
known \citep{bottcher2006reconstruction}; what matters here is that it applies
to the occupancy and drift channels unequally.  The
occupancy channel never forms a difference, so the same error cannot enter
twice.  It only blurs the heading density, convolving it with the known error
law, and a known blur is a deterministic smoothing that can be undone inside the
likelihood.  Sec.~\ref{sec:validation} verifies both statements and
Note~S1 gives the algebra.

\subsubsection{The detectability law}

Write $z_n$ for the asymptotic signal-to-noise ratio of harmonic $n$, evaluated
at the true coefficient,
$z_n=\Gamma_n/\mathrm{SE}(\hat\Gamma_n)$.  The observed coefficient divided by
its cluster-robust standard error is the corresponding finite-sample statistic.  Its
denominator is fixed by Eq.~\eqref{eq:variance},
\[
  \mathrm{SE}(\hat\Gamma_n)
  =\sqrt{\frac{4D}{T\,n^2\,\mathbb E|F_n|^2}}
  =\frac{2}{n}\sqrt{\frac{D}{T\,\mathbb E|F_n|^2}},
\]
and its numerator is the coefficient itself, with $\Gamma_n\propto K_n$, so
\[
  z_n\;\propto\;\sqrt{\frac{T}{D}}\;n\,K_n\,\sqrt{\mathbb E|F_n|^2},
\]
the constant of proportionality being the same at every harmonic.  Now
substitute the factorization of Eq.~\eqref{eq:factorization}.  The explicit
$n$ cancels against the $1/n$ that the neural form factor carries,
\[
  n\,K_n=J_{\mathrm{int}}\,n\,c_n(\sigma)\,M_n(W)
  =\frac{4J_{\mathrm{int}}}{\pi^{2}}\,e^{-n^2\sigma^2/2}\sin\!\Big(\frac{nW}{2}\Big),
\]
which is why no explicit $n$ survives below.  Dividing by the same quantity at
$n=1$ leaves the significance attainable at harmonic $n$ relative to harmonic
$1$ in a fixed experiment,
\begin{equation}
  \boxed{\;
  \frac{z_n}{z_1}
  =\underbrace{e^{-(n^2-1)\sigma^2/2}}_{\text{sensory}}\;
   \underbrace{\frac{\sin(nW/2)}{\sin(W/2)}}_{\text{neural}}\;
   \underbrace{\sqrt{\frac{\mathbb E|F_n|^2}{\mathbb E|F_1|^2}}}_{\text{design}}\;}
  \label{eq:detectability}
\end{equation}
and, for fixed diffusion, kernel and design, the asymptotic observation time
needed to reach a given signal-to-noise ratio at harmonic $n$ exceeds that for
harmonic $1$ by the inverse square of this ratio.  The
sensory factor is Gaussian in $n^2$ and dominates.  In a spectrally flat
design at $\sigma=0.7$~rad, a third harmonic requires $13$ to $50$ times more
data than a first across bump widths $W=60$--$90^\circ$, and the requirement
diverges as $W$ approaches $120^\circ$, where the neural filter itself has a
zero at $n=3$.  At $\sigma=0.9$~rad the sensory factor raises the cost to
$163$--$652$ times (\figpanel{fig:selection}{e}, left).  This is a quantitative
prediction of the theory, and it accounts for a striking regularity in the
literature: trajectory-based interaction inference rarely resolves angular
structure beyond the first harmonic
\citep{katz2011inferring,herbertread2011inferring,calovi2018disentangling}.  It
is tempting to read that as evidence that angular responses are simple.
Eq.~\eqref{eq:detectability} says instead that it is what a broadly tuned
sensory front end must produce, and it tells an experimenter exactly how much
observation, or how much leverage, is required to see past it.

\subsection{The spectral origin of decision bifurcations}
\label{sec:bifurcation}

Higher harmonics can be difficult to measure yet strongly influence the
decision transition.  Place two identical cues at $\pm\Delta/2$ and measure heading
from their bisector, so the bisector is $\phi=0$.  The landscape is then
$H_\Delta(\phi)=-\sum_n K_nF_n(\Delta)\cos(n\phi)$, with $F_n$ from
Eq.~\eqref{eq:pair-sf}.  Differentiating once,
\[
  H_\Delta'(\phi)=\sum_{n\ge1}n\,K_n\,F_n(\Delta)\,\sin(n\phi),
\]
which vanishes at $\phi=0$ at every separation: the bisector is always a
stationary point, by the symmetry of the pair about it.  What changes with
$\Delta$ is whether it is a valley the animal settles into or a ridge it falls
off, and that is the sign of the second derivative there.  Differentiating
again brings down a second factor of $n$, and at $\phi=0$ every cosine equals
one, so
\begin{equation}
  C(\Delta)\equiv H_\Delta''(0)=\sum_{n\ge1}n^2K_n\,F_n(\Delta)
  =2\sum_{n\ge1}n^2K_n\cos\!\Big(\frac{n\Delta}{2}\Big),
  \label{eq:curvature}
\end{equation}
the second equality substituting $F_n(\Delta)=2\cos(n\Delta/2)$.  The bisector
is locally stable exactly when $C(\Delta)>0$, and the critical separation
$\Delta_c$ --- where compromise gives way to selection --- is where this
curvature changes sign, $C(\Delta_c)=0$
\citep{salahshourcouzin2026harmonic}.  The factor $n^2$ weights each harmonic by
the square of its order, so the transition is decided by exactly the harmonics
that are hardest to fit.  Two consequences follow.

\paragraph{A first harmonic cannot create a local bifurcation.}
If $K_n=0$ for $n\ge2$ then $C(\Delta)=2K_1\cos(\Delta/2)>0$ for every
$\Delta<\pi$.  For a fixed symmetric pair with equal effective cue gains, the
bisector is therefore stable at every separation short of antipodal.  A local
compromise-to-selection instability at $\Delta_c<180^\circ$ requires
higher-harmonic structure; such transitions are documented by
\citet{sridhar2021geometry}.
This statement concerns the conditional landscape at fixed geometry: a
distribution pooled over frames with different cue gains or resultants can
become bimodal even when every conditional first-harmonic landscape is
unimodal, a distinction tested explicitly below.

\paragraph{The critical angle is a harmonic fingerprint.}
Truncating at one higher mode, $C(\Delta_c)=0$ gives
\begin{equation}
  \frac{K_h}{K_1}=-\frac{\cos(\Delta_c/2)}{h^2\cos(h\Delta_c/2)},
  \qquad h\in\{2,3,\dots\}.
  \label{eq:required-ratio}
\end{equation}
In the one-extra-harmonic comparison of Eq.~\eqref{eq:required-ratio}, the two
candidate harmonics behave very differently.  A transition near
$\Delta_c=100^\circ$ requires $K_2/K_1\approx0.93$ but only
$K_3/K_1\approx0.082$.  As $\Delta_c\to90^\circ$ the second-harmonic
requirement diverges, because $\cos\Delta_c\to0$, so a second harmonic of any
size cannot place the transition there.  A third harmonic of a few per cent
suffices (\figpanel{fig:selection}{e}, right).  Within that restricted comparison,
critical angles in the $90$--$110^\circ$ range are more compatible with a small
third harmonic than with a second harmonic alone.  By
Eq.~\eqref{eq:detectability} that same ratio of $0.082$ at
$n=3$ leaves $z_3/z_1\approx0.25$ even in a spectrally flat design, so reaching
the same significance takes sixteen times the observation.  A real experiment
need not attain this precision: the squared ratio of the largest attained
first- and third-harmonic signal-to-noise in these recordings is $69$ (S1).

The bifurcation and the spectrum are thus two views of one object, and they
have complementary statistical properties.  Higher coefficients can be difficult
to estimate because they are small, whereas the location of a bifurcation can
be highly sensitive to them.  Sec.~\ref{sec:results} uses each to check the other.

\subsection{Seven archives, and what the estimator recovers}

\subsubsection{Seven archives, one instrument}
\label{sec:roadmap}

We now test the theory using both simulations and public datasets.  We check
the estimator first on simulations with known coefficients
(Sec.~\ref{sec:validation}), and then on seven archives: tethered
\emph{Drosophila} steering among two or three virtual targets in closed loop
\citep{sridhar2021geometry,godm2021data}, the designed array whose geometry
tunes the filter; ellipsoid-body activity recorded in flight under panoramas
of one-, two- and four-fold symmetry \citep{kim2019generation}; two-photon
compass recordings carrying the cue direction in the fly's own frame, with cue
contrast varied within fly in a subset
\citep{basnak2025multimodal,basnak2024dandi}; rummy-nose tetra near the arena
wall, alone and in groups \citep{xue2023light,xue2023data}; and three tests of
reach --- larval-zebrafish phototaxis \citep{karpenko2020light}, mouse head
direction \citep{ajabi2023population}, and two-beam phototaxis in a
single-celled alga \citep{raikwar2025phototactic}.  Table~S2 lists every
claim, the test applied to it and the outcome.

Cue geometry filters the behavioural response as the selection rules require:
with two targets and again with three, the second harmonic reverses sign where
that array annihilates it (Secs.~\ref{sec:h2} and~\ref{sec:threecue}).  The
same rules govern the compass, where a scene's programmed symmetry pushes down
the moments it forbids and shifts every prespecified within-fly discriminator
in the predicted direction in every fly (Sec.~\ref{sec:compassrule}).  The
shape that sets the spectrum is measurable too: the compass bump narrows as cue
contrast rises, and its width predicts the harmonics it should carry
(Sec.~\ref{sec:contrast}).  The fish's response to a wall is strongly
even-harmonic, so a simple repulsion rule misses most of it
(Sec.~\ref{sec:tetra}).  Universality does not follow.  The fitted
second-harmonic coefficient does not carry from one cue arrangement to the next
(Sec.~\ref{sec:transfer}), and the pooled fly transition does not by itself
overturn vector averaging (Sec.~\ref{sec:h1-transition}).

\subsubsection{The estimator does what the theory says}
\label{sec:validation}

We first checked the predictions of
Secs.~\ref{sec:diffraction}--\ref{sec:channels} on simulated trajectories:
cues placed at bearings we choose, a heading integrated from the theory's own
stochastic equation, and coefficients we set, so the truth is known.
Note~S1 gives each design and each check in full.
Selection rules hold exactly: on a three-cue $120^\circ$ array the estimator
reports the annihilated $\kappa_1$ and $\kappa_2$ as unidentifiable rather
than as zero, and recovers $\kappa_3$ at $0.163$ against a true $0.167$.  Both
channels recover their targets with standard deviations that agree to within
$4\%$ and sit within $1.0$--$1.35\times$ of the asymptotic prediction
Eq.~\eqref{eq:variance}, so the occupancy channel gives up essentially nothing
by discarding the increments, as Proposition~\ref{prop:equivalence} requires.
The two channels diverge under measurement error, exactly as predicted:
at heading noise of $0.25$~rad the drift estimator's maximum relative error
reaches $863\%$ against the occupancy estimator's $49\%$.  The consistency
ratio $R_n$ (Eq.~\eqref{eq:fdt}) sits at one when the two channels share the
same aligned gradient dynamics and departs only where an out-of-phase term is
injected.  In the moving-observer validation, the closed loop between
heading and cue geometry did not manufacture appreciable higher harmonics:
under a purely first-harmonic rule the estimator returns $\kappa_2=-0.008$ against a first
harmonic of $0.822$.

\subsection{The behavioural spectrum and its selection rules}

\subsubsection{Higher harmonics shift the predicted transition toward the data}
\label{sec:h1-transition}

\begin{figure}[p]
\centering
\includegraphics[width=\textwidth]{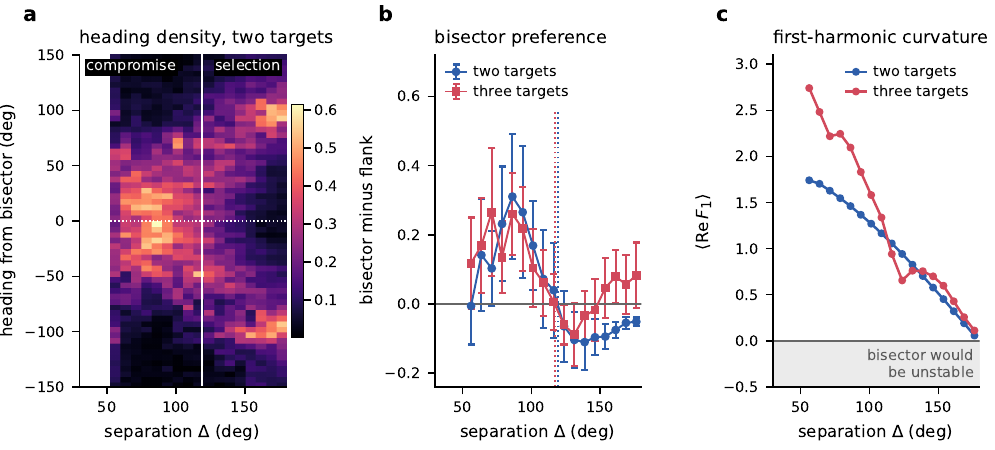}
\caption{
\textbf{The observed compromise-to-selection transition.}
\emph{Drosophila} in virtual reality \citep{sridhar2021geometry,godm2021data}.
\textbf{a}, Heading density relative to the bisector of two virtual targets,
as their angular separation sweeps.  At small separations the density holds a
single peak on the bisector; by large separations that peak has split toward
the targets.  The white line marks the measured transition at $119.1^\circ$.
\textbf{b}, The same change without a density map: the density averaged within
$15^\circ$ of the bisector minus the combined flank density $30$--$60^\circ$
away, in the two- and three-target cohorts.  This model-free contrast changes
sign at the observed transitions (dotted lines, $119.1^\circ$ and
$116.9^\circ$).  It measures whether heading prefers the bisector or its
flanks; it is not the local curvature of the landscape.
\textbf{c}, The measured first-harmonic structure factor of the realised cue
geometry.  It remains positive throughout both sweeps, so
Eq.~\eqref{eq:curvature} predicts a stable bisector for the equal-gain,
fixed-geometry first-harmonic landscape; the shaded region is where that
landscape would instead predict instability.  A first harmonic alone therefore
does not produce the split seen in \textbf{a} and \textbf{b}.  This
geometry-only diagnostic does not exclude a split after conditional densities
with distance-dependent gains are pooled; S3 tests that fitted model directly.
Data: $30$ two-target recordings holding $319\,647$ retained frames and $30$
three-target recordings holding $319\,502$. Rows are
thinned to at most one per $50$~ms of acquisition time and the first $0.5$~s
after each display switch is dropped; target bearings and distances are
reconstructed from the archive's experiment and project databases.  All three
panels bin separation at $7.5^\circ$ between $30^\circ$ and $180^\circ$ and
drop any bin holding fewer than $1500$ frames, which removes the three bins
below $52.5^\circ$ and leaves $312\,386$ and $316\,996$ frames; \textbf{a}
additionally bins heading at $5^\circ$.  The unit of replication is the
recording: each point in \textbf{b} is the mean over the recordings
contributing at least $200$ frames to that bin, with a $95\%$ interval from
their spread.  No fitting or resampling enters any panel.  Full
method in Note~S3.}
\label{fig:godm}
\end{figure}

We now turn to the fly recordings.  Measuring heading from the bisector of the
two targets and binning by their angular separation gives a model-free picture
of the decision (\figpanel{fig:godm}{a}): at small separations the heading
density holds a single peak on the bisector, and at large separations that
peak splits toward the targets.  We quantify this with a
\emph{centre-versus-flank density contrast} (\figpanel{fig:godm}{b}): the
probability density averaged within $15^\circ$ of the bisector minus the
combined-flank density averaged $30$--$60^\circ$ away.
The name is deliberate: this is a coarse, nonlocal measure of whether heading
prefers the bisector or its flanks, not the local curvature of the landscape,
and its sign can in principle differ from the sign of that curvature when
remote modes carry the mass.  Restricting to \emph{downward} crossings above
$60^\circ$ --- crossings at which the contrast falls through zero as the
separation widens --- the contrast changes sign at $\Delta=119.1^\circ$ in the
two-target cohort and $116.9^\circ$ in the three-target cohort, returning to
positive at $148.3^\circ$.  The window and direction were chosen after
inspection of the contrast curve but before comparison with any harmonic
prediction.  The contrast is
not monotone across the whole sweep: it rises before it falls, and the
unrestricted series also has an early \emph{upward} crossing at $56.5^\circ$,
driven by a single weakly negative bin at $56.25^\circ$ ($-0.006\pm0.057$),
which the window excludes on both counts.  No harmonic model enters
any of these numbers.

For a fixed geometry, any weighted sum of first-harmonic cue terms reduces to
one sinusoid directed toward their vector resultant.  With equal cue weights,
$\langle\mathrm{Re}\,F_1\rangle$ remains positive throughout the sweep: its
minimum is $+0.057$ in the two-target cohort and $+0.112$ in the three-target
cohort (\figpanel{fig:godm}{c}), so the curvature of the mean equal-gain
first-harmonic landscape does not reverse.  The density shown, however, pools
frames recorded at different geometries.  Because cue gain varies with distance, a mixture of unimodal
first-harmonic conditional densities can itself become flank-dominated.

We therefore refitted the distance-resolved first-harmonic model as its own
maximum-likelihood model and passed its conditional density through exactly
the centre-versus-flank statistic above (S3).  In the two-target cohort, it
predicts a crossing at $134.4^\circ$ (recording-cluster sandwich range
$[131.1^\circ,137.8^\circ]$), later than the observed $119.1^\circ$.  The full
four-harmonic model moves the prediction to $110.3^\circ$
$[95.1^\circ,124.8^\circ]$; retaining the laboratory-frame block moves the two
predictions to $143.9^\circ$ and $114.0^\circ$.  The higher modes therefore
bring the two-target crossing toward the data, although the cue-only
curve-wide RMSE changes only from $0.073$ to $0.072$.  In the three-target
cohort the refitted first-harmonic model has no crossing, while the cue-only
four-harmonic point curve crosses only at $175.7^\circ$ and the version with
the laboratory block has no point crossing, against $116.9^\circ$ in the
data.  Thus the pooled transition does not categorically falsify vector
averaging, and the fitted spectrum does not reproduce the three-target
transition.

\subsubsection{The second harmonic, measured through its selection rule}
\label{sec:h2}

The two-target design is a tunable harmonic filter, whose geometry-dependent
effect provides evidence for higher-harmonic structure.  We fit an
unconstrained landscape $\log p(\phi)=\sum_n\lambda_n\cos n\phi$ separately in
each separation bin.  The theory predicts
$\lambda_n(\Delta)=\kappa_n\langle\mathrm{Re}\,F_n\rangle(\Delta)$
(Eq.~\eqref{eq:occupancy}): one constant $\kappa_n$ per harmonic, the same in
every bin.

\begin{figure}[p]
\centering
\includegraphics[width=\textwidth]{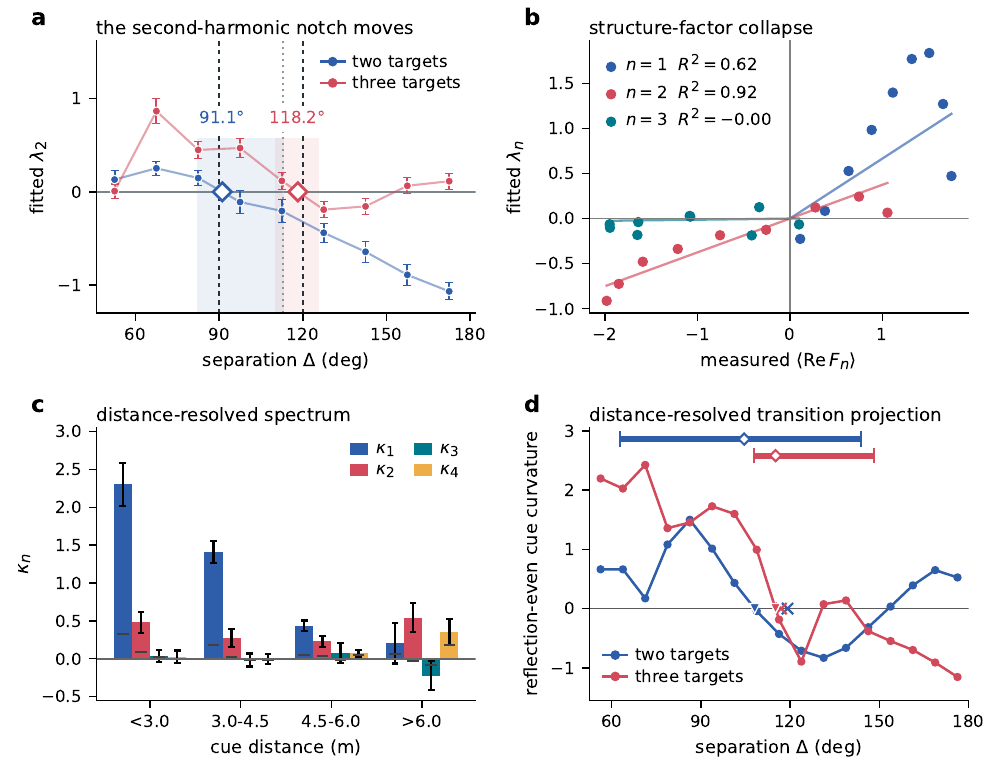}
\caption{
\textbf{The second harmonic follows cue geometry, and the distance-resolved
spectrum predicts an even-curvature reversal of uncertain location.}
\emph{Drosophila} in virtual reality \citep{sridhar2021geometry,godm2021data}.
\textbf{a}, The second-harmonic coefficient $\lambda_2$, fitted independently
in each separation bin, for two targets (blue) and three (red).  Each cohort
crosses zero near the location predicted by its cue geometry: the pair at
$91.1^\circ$ against a predicted $90^\circ$, the triplet at $118.2^\circ$
against a nominal $120^\circ$ (dashed) or the realised $113.0^\circ$ (dotted).
Diamonds mark the crossings, the shading their $95\%$ intervals,
$[81.9^\circ,113.2^\circ]$ and $[110.2^\circ,125.8^\circ]$.  Lines only join
separate bin fits; no constant-coefficient collapse is assumed for the
triplet.
\textbf{b}, Two-target coefficients against measured
$\langle\mathrm{Re}\,F_n\rangle$, $n=1,2,3$: agreement is strongest at
$n=2$; $n=3$ does not collapse.  Lines are through-origin fits; $R^2$ is
inverse-variance-weighted and centred.
\textbf{c}, The two-target spectrum resolved by cue distance: the
first-harmonic response weakens with distance while the second stays clear of
zero in every stratum; dashes show one geometry reassignment.
\textbf{d}, The cue-only curvature at the bisector, reconstructed within each
distance stratum through $n=4$ then averaged over the realised geometry.
Curves are the reflection-even projection about that bisector, not the
curvature of a guaranteed stationary point.  Triangles mark its first downward
zero ($108.1^\circ$, $115.1^\circ$), crosses the independent
centre-minus-flank reversal in the same recordings ($119.1^\circ$,
$116.9^\circ$).  The reversal is firm in direction and weak in location: bars
and open diamonds are conditional $95\%$ ranges and medians, a crossing is
defined in $91.9\%$ and $99.9\%$ of draws, and more than one occurs in
$33.9\%$ and $56.2\%$.  The reflection-odd slope at the plotted crossing is
$0.003$ and $-0.119$, so the unsymmetrised three-target landscape need not be
stationary at the bisector.
Archive and preprocessing as in Fig.~\ref{fig:godm};
the recording is the unit, $30$ per cohort.  \textbf{a},\textbf{b} use the
nine $15^\circ$ bins holding at least $5000$ rows ($313\,596$ and $318\,470$
frames); \textbf{a} is fitted with a laboratory-frame block and \textbf{b}
without.  Bars are joint recording-cluster
standard errors, crossing intervals from $100{,}000$ Rademacher cluster-score draws
under the first-downward-crossing rule
(Methods).  \textbf{c} strata: $0$--$3$, $3$--$4.5$, $4.5$--$6$, ${>}6$~m.
\textbf{d} uses $7.5^\circ$ bins, frame weighting, at least $200$ rows per
recording-bin, $312\,386$ and $316\,996$ frames, and $50{,}000$ sandwich draws. Full
method in Notes~S2 and S3.}
\label{fig:bifurcation}
\end{figure}

The second harmonic follows this prediction closely
(\figpanel{fig:bifurcation}{a,b}).  For each frame, we calculated $F_n$ directly from
the reconstructed target bearings and then averaged $\mathrm{Re}\,F_n$ within
each separation bin; this geometric quantity is not fitted from behaviour.  For
two identical targets, $F_2=2\cos\Delta$ (Eq.~\eqref{eq:pair-sf}), which
vanishes at $\Delta=90^\circ$, and the fitted behavioural coefficient
$\lambda_2$ changes sign at $90.0^\circ$.  Regressing $\lambda_2$ on the
measured structure factor through the origin collapses the nine bins onto one
line, giving $\kappa_2=0.375\pm0.027$ (conditional standard error) and an inverse-variance-weighted
$R^2=0.917$.  The first harmonic also collapses reasonably well
($\kappa_1=0.666\pm0.028$, $R^2=0.620$); its remaining variation is consistent
with the strong distance dependence resolved below.  In contrast, the third
harmonic does not track its predicted structure factor ($R^2=-0.003$), and its
collapse slope is consistent with zero ($\kappa_3=0.012\pm0.023$).  The fourth
harmonic has $R^2=0.309$, but is not stable enough across geometry to support a
common cue-locked coefficient.

An anisotropic arena --- a brighter wall, a door or another visible gradient
--- produces heading structure fixed in the laboratory frame that can be
mistaken for a cue-locked response.  In these data, the laboratory bias is
specifically twofold, so its $\cos 2\Phi$ and $\sin 2\Phi$ components can
contaminate the estimated second harmonic.  We therefore refitted each
separation bin with both components simultaneously:
\[
\log p_b(\Phi_i)
=\sum_n\lambda_{n,b}\cos\!\big[n(\Phi_i-B_i)\big]
+\sum_m\left[a_{m,b}\cos(m\Phi_i)+b_{m,b}\sin(m\Phi_i)\right]
-\log Z_{ib},
\]
where $B_i$ is the cue bisector.  The first sum rotates with the cue geometry;
the second remains fixed relative to the room.  The cue-locked second harmonic
survives this control.  Comparing like with like, the two arms are unweighted
through-origin re-slopes of the same per-bin coefficients: $\kappa_2=0.427$
with the laboratory block and $0.348$ without it, while the sign change moves
only from $90.0^\circ$ to $91.1^\circ$.  This convention differs from the
inverse-variance-weighted estimate $0.375$ above; the corresponding weighted
comparison is $0.433\pm0.027$ against $0.375\pm0.027$ (Supplementary Note~S2).
The anisotropy absorbed by the block is real and specifically second-harmonic:
blocks at other orders have no effect, and randomly rotating each recording
abolishes the laboratory-frame effect while preserving every cue-relative
quantity.  Thus, the control shows that the second harmonic is not an artifact
of arena anisotropy, although it does not establish that its magnitude is more
precisely known (Supplementary Note~S2).

Separation bins reuse the same animals, so bin-level standard errors are not
independent and a regression across bins understates the uncertainty.  A
whole-recording bootstrap gives $\kappa_2=0.424$ with a $95\%$ interval of
$[0.341,\,0.525]$.  For the notch, we jointly propagated the recording-cluster
score covariance of all nine bin fits with $100{,}000$ Rademacher multiplier
draws, applying the first positive-to-nonpositive crossing as separation
increases (Methods).  The point crossing is $91.1^\circ$ and the median and
$95\%$ interval are $91.1^\circ$ $[81.9^\circ,\,113.2^\circ]$; a crossing is
defined in every draw (\figpanel{fig:selection}{c};
\figpanel{fig:bifurcation}{a}).  The coefficient is well separated from zero, whereas
the notch is not localised to a degree and is therefore quoted with its
interval wherever precision matters.

The collapse also needs a null that preserves the nominal cue layout.  Although
the targets are fixed in the arena, each fly's movement generates a different
time series of target bearings and angular separations.  In a full
within-layout permutation diagnostic, the observed unweighted
second-harmonic collapse ($R^2=0.863$) was not unusual enough to reject this
conditional null ($p=0.092$); neither was the maximum statistic over
$n=1,\ldots,4$ ($p=0.113$).  A fixed-point-free sensitivity gave a smaller
second-harmonic tail but does not form the complete permutation group.  The
collapse is therefore descriptive support rather than an independent
significance test; the directly fitted coefficient and its geometry-locked
sign reversal provide the inferential evidence (Supplementary Note~S3).

To separate distance-dependent response strength from spectral shape, we fitted
the target contributions jointly, with separate harmonic coefficients for four
fly-to-target distance strata: $0$--$3$, $3$--$4.5$, $4.5$--$6$ and $>6$~m
(\figpanel{fig:bifurcation}{c}).  The first-harmonic coefficient decreases across these
strata, from $\kappa_1=2.31$ to $1.41$, $0.44$ and $0.20$, showing that the
overall directional response weakens farther from the targets.  The second
harmonic remains distinguishable from zero in every stratum, with
$\kappa_2=0.479$, $0.275$, $0.228$ and $0.540$ and corresponding $t=7.0$,
$4.4$, $6.1$ and $5.6$.  In the two nearest strata, where $\kappa_1$ is most
reliably estimated, the normalized contributions are similar:
$\kappa_2/\kappa_1=0.21$ and $0.20$.  Thus, the response weakens with distance,
while its relative second-harmonic contribution is similar over the two
best-resolved strata.

Two sensitivity analyses tested the second harmonic against specified effects
of heading-observation error and the arena boundary.  Assuming wrapped-normal
heading errors of $0.05$ or $0.15$~rad changes $\kappa_2/\kappa_1$ by less than
$7\%$.  Restricting the fit to frames within four fifths of the arena radius
changes the same ratio from $0.208$ to $0.268$ in the nearest distance stratum
and from $0.195$ to $0.193$ in the next; all remain within the corresponding
reported approximate cluster-robust intervals from the full-data fit.  Neither
specified correction removes the second harmonic, although these checks cannot
exclude every possible measurement bias (Supplementary Note~S2).

\subsubsection{A three-cue geometry tests the rule at a different angle}
\label{sec:threecue}

The three-cue cohort tests the same selection rule against its own geometry.
Its three targets sit at $-\Delta/2$, $0$ and $+\Delta/2$, so
$\mathrm{Re}\,F_2=1+2\cos\Delta$, which vanishes at $120^\circ$ --- not at the
$90^\circ$ of a symmetric pair.  The two-cue notch
$180^\circ/n$ therefore does not apply to this geometry.

The central three-target result is the downward second-harmonic sign reversal
predicted by its cue geometry.  We locate it with the rule fixed for the
two-target cohort: the first crossing of the fitted $\lambda_2(\Delta)$ from
positive to negative as separation increases, linearly interpolated between
adjacent bin midpoints.  With the laboratory-frame block included, the point
crossing is $118.2^\circ$; the same joint cluster-score propagation used above
gives a median of $118.2^\circ$ and a $95\%$ interval of
$[110.2^\circ,\,125.8^\circ]$, with a crossing defined in $99.6\%$ of draws
(\figpanel{fig:bifurcation}{a}).
This interval contains both the
nominal $120^\circ$ prediction and the realized structure-factor zero at
$113.0^\circ$: because the flies translate through the arena, the bearings
they actually sample shift the geometric zero about $7^\circ$ below that of
the drawn design (Methods).  The location is stable to analysis choices: re-binning
changes it by only $0.9^\circ$, and an estimator binned by structure factor
rather than separation places it at $117.0^\circ$.  Two qualifications remain
(S3): the crossing rule was chosen after inspecting the point curves, although
before comparison with the predicted angles; and the notch belongs to the
pooled sweep, whose mixture of nominal cue arrangements is determined by the
trajectories, rather than to any single arrangement.

This local notch does not establish a full-sweep collapse.  The collapse makes
the stronger claim that all nine bins satisfy
$\lambda_2=\kappa_2\langle\mathrm{Re}\,F_2\rangle$ with one constant
$\kappa_2$.  They do not.  The primary inverse-variance-weighted fit gives
$\kappa_2=0.069$ and $R^2=-0.049$; the corresponding unweighted statistic is
negative, $R^2=-0.101$, and the maximum over all four harmonics is
unexceptional under the within-layout reassignment diagnostic ($p=0.984$;
S3).  The third harmonic is detectable in this cohort
($\kappa_3=0.073$, $t=3.2$), unlike in the two-target cohort
($\kappa_3=0.012$, $t=0.5$).  A separate joint cue- and laboratory-frame
bootstrap nevertheless gives a reliably positive pooled second-harmonic slope,
$\kappa_2=0.135$ $[0.093,\,0.194]$.  This is not evidence of a successful
collapse: an average slope can be nonzero even when one constant coefficient
does not describe the full sweep.

The failure arises from design mixing.  The three-target cohort contains three
nominal cue arrangements rather than one arrangement varied continuously in
separation.  As a fly moves, its perceived cue separation changes, so each
separation bin contains a different mixture of the three arrangements, in
proportions determined by the trajectories.  In the laboratory-controlled
unweighted analysis, fitting within nominal arrangements reduces the pooled
slope from $0.138$ to $0.005$; the identical conditioning leaves the two-target
estimate essentially unchanged ($0.427$ to $0.429$; S3).  The apparent
three-target slope is therefore driven mainly by changing arrangement
composition, not by a common within-arrangement coefficient.  The conclusion
is limited but clear: the pooled curve contains a stable, geometry-predicted
notch, but the data do not support one constant $\kappa_2$ across the
three-target sweep.

\subsection{What the spectrum predicts, and what it does not carry}

\subsubsection{The two channels agree on the spectrum}
\label{sec:fdt}

Running the drift channel on the same recordings gives an independent check
that does not share the occupancy channel's assumptions.  The shape comparison
needs no diffusion constant, because $D$ cancels from the normalized spectrum
(Eq.~\eqref{eq:occupancy}).
In the two-target cohort the occupancy channel gives
$\kappa_n/\kappa_1=(1,0.36,0.09,0.09)$ and the drift channel gives
$\Gamma_n/\Gamma_1=(1,0.30,0.04,-0.02)$: the second harmonic is present in
both, at close to the same relative size, and the increment regression finds
it at $t=2.5,5.1,6.4$ across the three well-determined distance strata.  The
second-harmonic measurement is therefore not an artefact of the occupancy
statistic.

The absolute ratio $\Gamma_n/(D\kappa_n)$, which theory puts at one, falls
short, by about a factor of two at the well-determined harmonics.  The
shortfall may partly reflect the one-frame increments: successive heading
changes are positively correlated, and the increment estimate
of $D$ is $32$--$42\%$ below the measured value.  The quadrature component omitted by the instantaneous cue-aligned model is small and, at
the first harmonic, carries the sign a response delay produces.  Note~S1
gives the full comparison, and with it the detectability accounting:
the squared ratios of the largest attained signal-to-noise across distance
strata are $7.1$ for the first versus second harmonic and $69$ for the first
versus third.  These describe attained detectability, not Fisher information.

\subsubsection{The heading dynamics, measured and cross-checked}
\label{sec:timescales}

Everything so far reads the experiment through Eq.~\eqref{eq:occupancy},
which is a stationary law: it gives the heading distribution after it has
equilibrated at a fixed cue geometry.  But the geometry is never fixed: the
animal's own motion sweeps the bearings continuously.  Whether the stationary
reading is legitimate comes down to two
time-scales --- how fast the heading distribution relaxes, and how fast the
geometry changes --- and both can be measured from the same recordings that
gave the spectrum.

Two dynamical quantities were estimated directly from the recordings without
using the stationary-spectrum fit.  The rotational diffusion coefficient $D$
in Eq.~\eqref{eq:sde} was obtained from the short-lag growth of the
mean-squared heading change, to which the landscape contributes only at second
order in the lag.  We obtained $D=0.287$~rad$^2\,$s$^{-1}$
$[0.211,\,0.373]$ for the two-target cohort and
$0.223$~rad$^2\,$s$^{-1}$ $[0.166,\,0.279]$ for the three-target cohort,
with $95\%$ intervals from a whole-recording bootstrap over the $30$
recordings in each cohort.  The mode-$1$ relaxation time $\tau_1$ was estimated
separately from the autocorrelation of heading relative to the cue bisector.
After subtracting its stationary plateau and normalising the remaining decay,
we fitted an exponential over the predeclared amplitude range
$0.65$--$0.95$.  This gave $\tau_1=3.21$~s $[2.63,\,4.18]$ and
$4.04$~s $[3.45,\,4.76]$, respectively (S4).

These measurements provide an out-of-fit dynamical test of the generative
model.  The stationary spectrum $\kappa_n$ was estimated from the heading
density using Eq.~\eqref{eq:occupancy}, without using temporal correlations.
We combined the distance-resolved spectrum with the separately measured $D$
in Eq.~\eqref{eq:sde}, solved the stationary Fokker--Planck dynamics at the
mean measured structure factors in each $15^\circ$ separation stratum, pooled
the predicted correlations by the observed frame counts, and applied the same
exponential fitting rule.  This stratified approximation gave
$\tau_1=3.50$~s for the two-target cohort and $3.54$~s for the three-target
cohort.  The measured times were $8\%$ below and $14\%$ above these
predictions, respectively, and both predictions lie inside the measured
$95\%$ intervals, with no additional fitted parameter.  These
values are close to the free-diffusion scales $1/D=3.48$ and $4.49$~s.  For a
pure first-harmonic landscape, the leading correction to the mode-$1$ rate is
second order in the coupling, but a multi-harmonic spectrum need not obey that
simplification; the reported predictions therefore use the full stationary
Fokker--Planck solution within each stratum (S1, S4).  The comparison links diffusion measured
from sub-second displacements to decorrelation over seconds and tests the
model's dynamical relation between the stationary landscape, noise amplitude
and relaxation.

With $\tau_1$ measured, the quasi-static approximation can be tested by asking
how far the cue geometry moves during one relaxation time.  In the experiment,
the fly was physically tethered, but its measured heading drove virtual forward
motion at $0.2\,\mathrm{m\,s^{-1}}$; the cue bearings therefore changed as if
the fly were translating past fixed targets.  This feature comes from the
closed-loop VR design, although the same problem arises for any moving
observer.  The median cue separation changes by $8.5^\circ$ and $11.8^\circ$
in the two cohorts --- between one half and four fifths of a $15^\circ$
separation bin --- whereas the cue bisector turns by only $2.1^\circ$ and
$2.9^\circ$.  This difference follows from the geometry: separation is the
difference between the two cue bearings, so their changes add, whereas the
bisector is their mean, so those changes largely cancel when the animal heads
near the bisector.  The stationary-occupancy approximation is therefore
reasonable for typical frames but less secure in the fast-sweeping tail (S4).

We tested whether the remaining cue motion biases the estimated first harmonic.
Simulation is needed because the true $\kappa_1$ is unknown in the experimental
data; by simulating a known $\kappa_1$, we can measure how much the stationary
estimator distorts it.  Two simulation approaches disagree even on the
direction of the bias.  In a fully simulated closed loop, the model animal
steers relative to the landscape and thereby changes its own cue bearings; this
makes the recovered first harmonic $35\%$ and $58\%$ too small at two operating
points.  In an open-loop replay, each animal's recorded cue sequence is retained
and only its heading is simulated; this instead makes the recovered first
harmonic $4.6\%$ and $8.2\%$ too large.

The two approaches produce similar cue-sweep rates but differ in whether the
simulated animal generates its own cue sequence through steering, as well as
in their geometry, boundary and initial-condition construction.  Their
opposite biases show that sweep speed alone is insufficient and are consistent
with a role for sensorimotor feedback.  However, the closed-loop simulation uses an idealized model
animal, whereas the replay breaks the real feedback loop; neither directly
identifies the bias in the experimental estimate.  We therefore apply no
correction: all reported spectra are the original fits, and both the magnitude
and direction of any residual bias remain open (S4).

\subsubsection{The distance-resolved spectrum predicts an even-curvature reversal}
\label{sec:prediction}

The harmonic coefficients and the centre-versus-flank contrast are different
summaries of the same recordings.  Their agreement is therefore an internal
consistency check, not an independent validation.  The comparison must also
retain cue distance: because response strength varies sharply with distance
(Sec.~\ref{sec:h2}), and distance covaries with separation, pooling the four
distance-block coefficients before evaluating Eq.~\eqref{eq:curvature} does
not give the curvature of the fitted model.

Let $D_r$ denote the four distance strata, $B_i$ the instantaneous cue bisector
in frame $i$, and
\[
F_{irn}=\sum_{j:\,d_{ij}\in D_r}w_{ij}e^{-\mathrm i n\Theta_{ij}}.
\]
For each $7.5^\circ$ separation bin $b$, we therefore evaluated
\[
C_b^{\rm even}=\frac{1}{N_b}\sum_{i\in b}\sum_{r=1}^{4}\sum_{n=1}^{N}
 n^2\widehat\kappa_{rn}\,\mathrm{Re}\!\left(F_{irn}e^{\mathrm i nB_i}\right).
\]
This is the second derivative at the bisector of the mean fitted cue
landscape, equivalently the curvature of its reflection-even component.
Positive curvature denotes a local minimum only if the companion slope is
zero.  Distance-dependent cue gains can make the two effective cue strengths
unequal, in which case the unsymmetrised landscape has nonzero slope at the
geometric bisector.  We therefore call a zero of
$C_b^{\rm even}$ an even-curvature reversal, not necessarily a literal
pitchfork bifurcation; the corresponding odd slope is reported in S3.

The first-harmonic even-curvature curve has no reversal over the measured sweep.  Adding
the second harmonic produces a point crossing at $121.5^\circ$ in the
two-target cohort and $161.3^\circ$ in the three-target cohort; adding the
third gives $121.5^\circ$ and $112.8^\circ$, and the full $n\leq4$ spectrum
gives $108.1^\circ$ and $115.1^\circ$
(\figpanel{fig:bifurcation}{d}).  The model-free centre-versus-flank crossings
are $119.1^\circ$ and $116.9^\circ$.  Thus higher harmonics generate an
even-curvature reversal.  Its four-harmonic point crossing lies numerically
near the observed contrast crossing, especially in the three-target cohort,
but these are different functionals and the inferred curvature location
remains sensitive to truncation.

Uncertainty reinforces that caution.  Joint propagation of the full
recording-cluster sandwich covariance gives conditional $95\%$ ranges of
$[62.8^\circ,143.9^\circ]$ and $[107.8^\circ,148.3^\circ]$ for the first
$n\leq4$ downward crossing; a crossing is defined in $91.9\%$ and $99.9\%$ of
draws, respectively.  Many propagated curves have more than one downward
crossing, so these are conditional first-crossing summaries rather than
regular confidence intervals.  Giving each eligible recording equal
geometric weight moves the point crossings only to $110.7^\circ$ and
$115.5^\circ$, whereas including the fitted laboratory-frame nuisance terms
moves them to $127.3^\circ$ and $112.4^\circ$.  A $100$-draw recording
bootstrap that refits all coefficients and recomputes both geometry and the
observed contrast gives the same qualitative conclusion (S3).

The defensible result is therefore narrower than a calibrated transition
probability: the measured higher harmonics produce a reflection-even
curvature reversal that the first harmonic does not, but these data do not
determine a unique critical angle or establish a symmetric bifurcation of the
full landscape.  The $n^2$ weighting in Eq.~\eqref{eq:curvature} makes
critical-angle calculations unusually sensitive to weak higher harmonics.

\subsubsection{The notch is harder to imitate than the coefficient}
\label{sec:discrimination}

\begin{figure}[p]
\centering
\includegraphics[width=\textwidth]{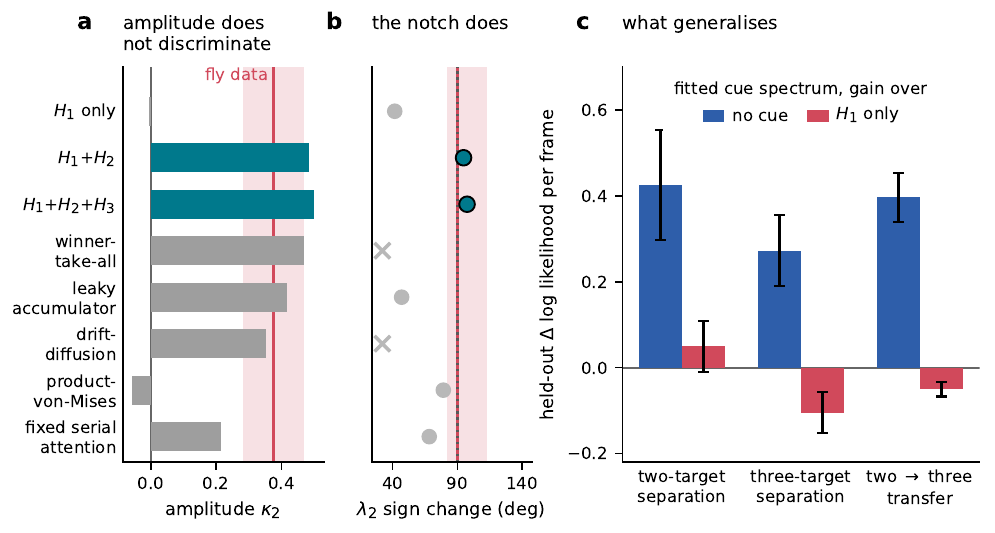}
\caption{
\textbf{Amplitude, notch and generalisation constrain the mechanism.}
Simulated turning rules, tested against \emph{Drosophila} in virtual reality
\citep{sridhar2021geometry,godm2021data}.
Eight fixed turning rules were run under a common idealised closed-loop
protocol, in which each simulated animal's own motion changed the target
bearings, and analysed with the same form of occupancy-and-collapse analysis
as the flies.
\textbf{a}, Fitted second-harmonic amplitude. Teal marks the two arms carrying an
explicit second harmonic, $H_1{+}H_2$ and $H_1{+}H_2{+}H_3$, and grey everything
else, including the first-harmonic-only control; the vertical line is the observed
fly value and
the shaded band $\pm25\%$ of it.  Amplitude does not identify a
mechanism: two alternatives, leaky accumulation and drift-diffusion, fall
inside the band and a third, winner-take-all, exceeds its upper edge by
$0.001$.  None of the three is parameterised as a harmonic landscape.  The
product-of-von-Mises rule returns a near-zero second harmonic, as expected
from its purely first-harmonic dependence on heading at fixed cue geometry.
\textbf{b}, Where each rule's per-separation coefficient $\lambda_2$ changes
sign.  The prediction is the $90^\circ$ notch of Eq.~\eqref{eq:pair-sf}
(dotted); the solid line is the flies' own fitted crossing, $90.0^\circ$,
which at this scale lies beneath it, and the shaded band is the
laboratory-controlled joint cluster-score interval for the flies,
$[81.9^\circ,\,113.2^\circ]$.  Rules with no sign change are drawn at the left
edge and marked $\times$.  Only the two explicit harmonic landscapes cross
inside the interval (outlined); none of the five fixed non-harmonic rules
does.  This comparison applies to the displayed parameterisations, not to
every member of the corresponding model families.
\textbf{c}, What generalises.  Each bar is the held-out gain in mean log
likelihood per frame of the fitted cue spectrum over a comparator, a cue-blind
null in blue and a first-harmonic-only model in red.  The first two groups fit
on the smaller separations of one cohort and score on the larger ones held
out; the third fits on the two-target recordings and scores on the three-target
ones.  The cue field
generalises strongly against the null, but the fitted higher harmonics add
no robust advantage over the first-harmonic model: they gain nothing in the
two-target extrapolation and reduce predictive accuracy in the other two comparisons.
\textbf{a},\textbf{b}: each arm is $1500$ simulated
trajectories of $2500$ steps at $\mathrm{d}t=0.02$~s, two cues with
separations drawn uniformly between $30^\circ$ and $180^\circ$, all eight arms from
one generator, with coefficients fitted in ten
$15^\circ$ separation bins.  The fly interval is the same
laboratory-controlled crossing interval as \figpanel{fig:bifurcation}{a}; the
fly amplitude is the collapse slope of \figpanel{fig:bifurcation}{b}, which
carries no laboratory-frame block.  \textbf{c}: the
separation tests fit below $105^\circ$ and score frames at $105^\circ$ and
above; the transfer test fits the two-target animals and scores the disjoint
three-target animals.  Bars are $95\%$ $t$ intervals over the $30$ recordings
per cohort across which each difference is paired.  This is the fixed-split
comparison; Note~S3 gives the stricter leave-one-recording-out cross-fitted
transfer quoted in the text, and Note~S5 the simulation equations, parameters
and decision rules.}
\label{fig:discrimination}
\end{figure}

A fitted second harmonic does not by itself identify a harmonic landscape.
We passed three harmonic controls and five non-harmonic turning rules through
a common fly-like closed loop and the same form of occupancy-and-collapse
analysis used for the fly data.  Winner-take-all, leaky accumulation and
drift-diffusion all produced second-harmonic coefficients close to the
observed value (\figpanel{fig:discrimination}{a}).  Coefficient magnitude is
therefore not mechanism-specific.

The geometry-locked sign change is more discriminating
(\figpanel{fig:discrimination}{b}; S5).  The harmonic landscapes collapsed onto
the predicted structure factor and crossed near the predicted $90^\circ$
notch.  Winner-take-all and drift-diffusion never crossed, whereas the leaky
accumulator crossed near $47^\circ$.  None of the five fixed non-harmonic rules
placed its crossing inside the laboratory-controlled empirical interval
$[81.9^\circ,\,113.2^\circ]$; the closest, a product-of-von-Mises rule at
$79.1^\circ$, also produced a near-zero second-harmonic amplitude.  This
comparison addresses the stated parameterisations, not every member of those
model families.

We then fitted ten partly overlapping model families on training recordings and
evaluated them on held-out recordings.  In this fixed five-fold comparison, the
null, first-harmonic, winner-take-all and boxcar families scored conditionally
worse than the harmonic model in both cohorts.  Serial attention remained
viable, while graded softmax attention tied the harmonic landscape in the
two-target cohort and had a small advantage in the three-target cohort.
Because the reported intervals resample held-out animal scores without
re-fitting folds, these rankings are conditional on the chosen split and are
not selection-adjusted.  Thus, the comparison favours a smooth, graded angular
response with higher-order structure, but does not uniquely identify the
harmonic form.  The
distinctive strength of the harmonic account is that cue geometry predicts
the notch location in advance (S5).

\subsubsection{What does not transfer}
\label{sec:transfer}

Three tests of extrapolation give a consistent and partly negative picture.
Training on separations below $105^\circ$ and predicting above it, the full
spectrum outperformed the no-cue null by $+0.425$ nats per frame in the two-target
cohort ($p=2\times10^{-7}$, paired over 30 recordings) and $+0.273$ in the
three-target cohort ($p=2.4\times10^{-7}$).  Transferring coefficients fitted
on the two-target cohort to the three-target cohort --- disjoint recordings and a
different structure factor --- outperformed the null by $+0.406$ nats per frame
$[0.351,\,0.460]$, positive in $30$ of $30$ recordings.  That is the strict
version of the comparison, with the laboratory-frame nuisance block estimated
only on training recordings, one recording left out at a time (Note~S3 documents why
the strict version scores \emph{higher} than the in-sample one).  The cue-linked response is strong, and its first-harmonic component
generalizes across geometry and across recordings (\figpanel{fig:discrimination}{c}).

The higher harmonics do not.  Against a comparator that keeps the fitted first
harmonic and sets the rest to zero, the full spectrum gained only $+0.050$
nats per frame in the two-target separation extrapolation ($p=0.098$) and
\emph{lost} $0.105$ in the three-target one ($p=1\times10^{-4}$) and $0.051$ in
the cross-geometry transfer ($p=7\times10^{-7}$).  Consistently, the
three-target cohort's own second harmonic is weak
($\kappa_2/\kappa_1=0.08$--$0.17$) and its collapse against the structure factor
fails ($R^2=-0.049$).

Within the two-target design, the second harmonic is present and reproducible
across the separation sweep, and higher harmonics are required for the
even-curvature reversal; across designs and recordings, however, the fitted
second-harmonic coefficient does not transfer.  The
three-target configuration is also the harder case: because the targets sit at
fixed arena positions while the animal moves, the nominal middle target is not
between the other two in a large fraction of frames.  The configuration is then
not mirror-symmetric in the egocentric frame, and a cosine-only landscape is
misspecified there.  Whether the failure to transfer is animal-to-animal
variation in $\kappa_2$, a limitation of the two-harmonic truncation, or a real
breakdown of the shared-spectrum assumption is not resolved by these data, and
Sec.~\ref{sec:discussion} states what would resolve it.

\subsection{The same rules in the neural compass}

\subsubsection{The selection rule, read out in the compass itself}
\label{sec:compassrule}

Every selection-rule measurement so far reads the landscape from behaviour.
The fly's compass offers a neural readout of the same prediction.  EPG
neurons hold a single bump of activity whose angle tracks the visual scene
\citep{seelig2015neural,kim2017ring,kim2019generation}.  The symmetry prediction follows directly:
when a scene has exact $p$-fold symmetry, rotating it by $360^\circ/p$ maps
it onto itself, so nothing the fly sees can distinguish the $p$ ways the bump
may lock to it.  If the stationary ensemble preserves that symmetry ---
equivalently, if the locking branches are sampled with equal probability ---
the offset distribution repeats every $360^\circ/p$, and its circular moments
$Z_m=\langle e^{\mathrm i m\delta}\rangle$ vanish unless $p$ divides $m$.
An individual trial may remain in one branch and break this ensemble symmetry.
This is
Proposition~\ref{prop:selection} imposed on the compass rather than on the
turning response, with the symmetry printed on the display and the offset a
reconstructed bump angle minus the recorded scene position.  In the empirical
test, exact symmetry refers to the deposited periodic $88$-column pattern
coordinate.  The physical panorama spans $330^\circ$ and has a fixed
$30^\circ$ gap, so the instantaneous visible image is only approximately
symmetric; the measured prediction is therefore selective suppression rather
than literal annihilation in every trial (S7).

\begin{figure}[p]
\centering
\includegraphics[width=0.91\textwidth]{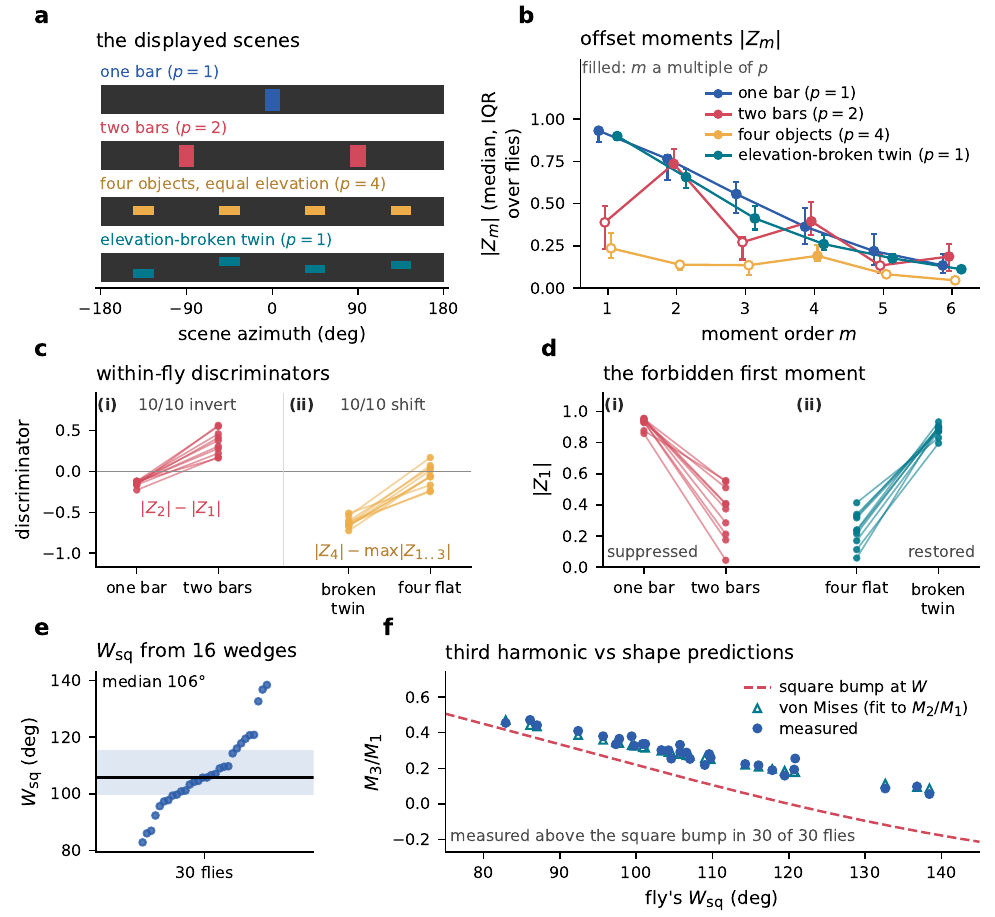}
\caption{
\textbf{The selection rule, read out in the fly's compass.}
\emph{Drosophila} compass imaging \citep{kim2019generation,kim2019data}.
\textbf{a}, The four scene classes, drawn schematically at their displayed
azimuths: one bar; two bars $180^\circ$ apart; four objects $90^\circ$ apart
at one elevation; and the elevation-broken twin, the same four azimuths at
four elevations.
\textbf{b}, Circular moments $|Z_m|$ of the bump-to-scene offset, fly-level
median with the interquartile range across flies.  Markers are filled at
multiples of the scene's symmetry order $p$ and open at forbidden orders.
Forbidden orders are selectively suppressed, and breaking the elevations
restores the one-bar form at unchanged azimuths.
\textbf{c}, The two prespecified within-fly discriminators, one line per fly:
\textbf{(i)}, $|Z_2|-|Z_1|$ between one bar and two bars; \textbf{(ii)},
$|Z_4|-\max|Z_{1..3}|$ between the broken twin and the flat scene.  Both move
as predicted in $10$ of $10$ animals.
\textbf{d}, The forbidden first moment alone: \textbf{(i)}, suppressed by the
programmed twofold symmetry; \textbf{(ii)}, restored by the elevation-broken
twin.  Suppression tracks symmetry, not object count.
\textbf{e}, The square-bump-equivalent width
$W_{\mathrm{sq}}=2\arccos(M_2/M_1)$ per fly, from the clean sixteen-wedge
ellipsoid-body ring; line, median $106^\circ$; band, interquartile range.
\textbf{f}, Measured $M_3/M_1$ against square-bump and von Mises predictions
matched through each fly's own $M_2/M_1$: all $30$ flies exceed the square
prediction, while the smooth profile stays consistent.
Data: $30$ flies, with the fly as the unit; \textbf{b} pools $30$,
$10$, $10$ and $20$ flies for the four scene classes; \textbf{c}\textbf{(i)}
and \textbf{d}\textbf{(i)} are one group of ten and \textbf{c}\textbf{(ii)} and
\textbf{d}\textbf{(ii)} another.  Preprocessing follows the authors'
offset pipeline and reproduces all $255$ usable of $260$ nominal trials in
those flies; claimed symmetries were checked against the periodic $88$-column
pattern stack, and the $330^\circ$ physical panorama is only approximately
symmetric (S7).  Flight-valid frames only; first $15$~s and trials under
$1000$ valid slices dropped; $200$ time shifts with a $10$~s margin per
$|Z_m|$.  \textbf{c},\textbf{d} are exact Wilcoxon tests on ten within-fly
pairs, floor $2/2^{10}$.  Full method in Note~S7.}
\label{fig:kim}
\end{figure}

We test it in public recordings of sixteen ellipsoid-body wedges imaged while
tethered flies flew in closed loop with panoramic scenes
\citep{kim2019generation,kim2019data}: thirty flies in three groups of ten
(\figpanel{fig:kim}{a}).  The displayed pattern stack is included in the data
deposit, allowing the symmetry of every scene to be verified directly.  A
single bar locks the offset, with moments decaying from a median $|Z_1|=0.93$,
whereas two identical bars $180^\circ$ apart suppress the odd moments and keep
the even ones (\figpanel{fig:kim}{b}).  For the two-bar contrast, the
prespecified discriminator $|Z_2|-|Z_1|$ inverts in $10$ of $10$ animals
(\figpanel{fig:kim}{c(i)}).  Four identical objects $90^\circ$ apart impose
four-fold symmetry, shifting the prespecified discriminator by $0.63$ toward
the fourth moment in $10$ of $10$ flies (\figpanel{fig:kim}{c(ii)}).  Both
statistics had their form and direction fixed by
Proposition~\ref{prop:selection} before the moments were examined.  The direct
first-moment comparisons show the same dependence: programmed twofold symmetry
suppresses $|Z_1|$ (\figpanel{fig:selection}{c}; \figpanel{fig:kim}{d(i)}), whereas companion scenes with the
same four azimuths at four different elevations break the symmetry and restore
first-moment locking, by $0.67$ within fly in $10$ of $10$ animals
(\figpanel{fig:kim}{d(ii)}).  The suppression therefore tracks symmetry, not
visual clutter.  The forbidden moments are suppressed but not eliminated:
some still exceed their per-trial null thresholds, mainly in a few flies,
which is why the fly-level discriminators rather
than the per-trial counts carry the inference (S7).  Each
prespecified contrast is ten within-fly pairs under an exact test whose smallest
achievable two-sided $p$ is the combinatorial floor $2^{-9}\approx0.002$, and
the contrasts sit exactly there, with every pair in the predicted direction.
Note~S7 gives the preprocessing certification, the per-trial null calibration,
and the sensitivity scans.

The same recordings also sample the compass bump cleanly as sixteen wedges
spanning the circle exactly once.  Writing $M_n$ for the $n$-th Fourier
coefficient of that bump profile, the square-bump-equivalent second-harmonic
width proxy $W_{\mathrm{sq}}=2\arccos(M_2/M_1)$ has a median of $106^\circ$
across the $30$ flies
(\figpanel{fig:kim}{e}).  The measured $M_3/M_1$ exceeds the
square-bump prediction at each fly's own $W_{\mathrm{sq}}$ in $30$ of $30$ flies
(\figpanel{fig:kim}{f}), while a smooth profile fitted only to $M_2/M_1$ remains
consistent with it (S7).  A smooth profile is consistent; a sharp-edged one is
not.

\subsubsection{The neural form factor and its measurement}
\label{sec:neural}

\begin{figure}[p]
\centering
\includegraphics[width=\textwidth]{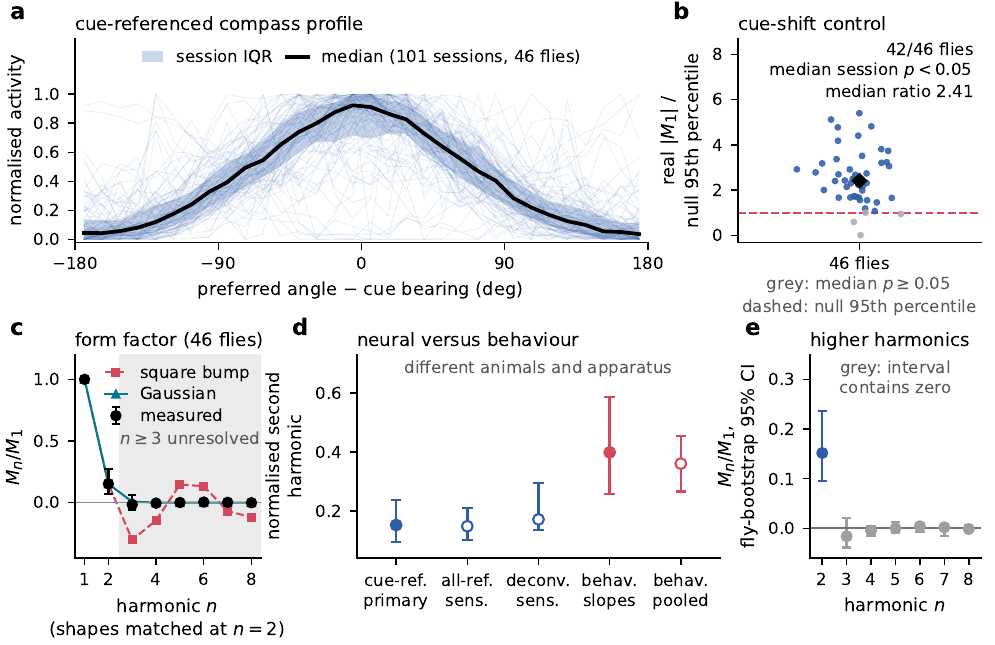}
\caption{
\textbf{The neural form factor, measured rather than assumed.}
\emph{Drosophila} compass imaging \citep{basnak2025multimodal,basnak2024dandi}.
\textbf{a}, Population activity against functional preferred angle minus cue
bearing.  Faint lines are individual sessions, the band is the interquartile
range across sessions and the heavy line the median.  Ring positions were
estimated on one half of each recording and the profile built from the other,
preventing tuning from being induced by sorting on the same data.
\textbf{b}, The cue-shift control, one point per fly: the fly's real $|M_1|$
divided by the $95$th percentile of its own null, built by circularly shifting
the cue series against the activity, which preserves the cue series while
destroying its correspondence with activity.  The dashed line is equality with
that null; the within-fly median tail fraction is below $0.05$ in $42$
of $46$ flies (grey otherwise), which summarises the session-level
fractions rather than being a fly-level permutation $p$-value.  It
tests first-harmonic cue locking, not the higher-order ratios in \textbf{c}.
\textbf{c}, The measured form factor, normalised to its first harmonic,
against the square and Gaussian activity bumps of Harmonic Theory
\citep{salahshourcouzin2026harmonic}, both matched
to the measured ratio $M_2/M_1=0.152$
alone, giving $W=162.5^\circ$ and $\sigma=1.12$~rad.  Only $n=1$ and $n=2$ are
resolved, and both models reproduce the matched ratio exactly, so the panel
shows a degeneracy rather than a discrimination.
\textbf{d}, The normalised second harmonic measured in the compass and in
behaviour, in different animals and apparatus.  The three neural entries are
the cue-referenced primary estimate, the all-reference sensitivity and the
upper-bound deconvolved sensitivity; the two behavioural entries are the
structure-factor collapse slope and the distance-pooled ratio.  The neural
ratios are the smaller: the raw neural interval does not overlap the
behavioural bracket and the deconvolved one does, so the panel constrains the
smoothing bridge without establishing it.
\textbf{e}, The neural ladder, $n=2$ to $8$.  Only $n=2$ excludes zero; every
higher order is consistent with it (grey), so this preparation resolves
nothing above the second harmonic.
Data: EPG recordings with cue bearing in the fly's own frame; $101$ eligible
sessions from
$46$ flies, each with at least $2000$ frames and $17$ to $53$ segmented
protocerebral-bridge regions.  Fourteen further sessions whose metadata
identifies angular zero with virtual heading rather than a unique cue are kept
only as a sensitivity analysis (S7).  The fly is the unit: sessions are
summarised componentwise within fly, and the intervals in \textbf{d} and
\textbf{e} are percentiles of $4000$ fly-level resamples, with the behavioural
entries carrying the corresponding recording-level
summaries (S3, S7). The null in \textbf{b} is $120$ shifts per session, so its $p$
floor is $1/121$. Full method in Note~S7.}
\label{fig:neural}
\end{figure}

The clean-ring analysis above constrains the bump's shape.  To connect the
neural profile quantitatively to behaviour, Harmonic Theory predicts that
$K_n$ inherits $M_n$, the $n$-th Fourier coefficient of the activity profile a
cue evokes on a ring of heading-tuned neurons.  We measured that profile in an
independent archive (\figpanel{fig:neural}{a}).

We used two-photon recordings of EPG compass neurons in the \emph{Drosophila}
protocerebral bridge with an angular cue bearing recorded in the fly's own
frame \citep{basnak2025multimodal,basnak2024dandi}.  The primary analysis contains the $101$
eligible sessions from $46$ flies whose immutable NWB metadata explicitly
identifies angular zero with the cue directly ahead; the $14$ further sessions
whose zero denotes virtual heading are retained as a sensitivity analysis
(S7).  The region index in these files
is a segmentation label rather than an angular coordinate, so the ring was
recovered empirically.  The recorded cue bearing is zero directly in front of
the fly, and each region is placed at a functional angle inferred from the
phase of its first-harmonic cue tuning, not at an anatomical coordinate.
Sorting regions by a quantity estimated from the same frames would
manufacture tuning out of noise, so preferred angles were estimated on one half
of each recording and the profile built from the other (S7).  A related
cosine-only split-half control circularly shifts the cue time series.  The
sessionwise empirical tail fraction is below $0.05$ in $83$ of $101$ sessions;
after repeated sessions are summarized by their within-fly median, it is below
$0.05$ in $42$ of $46$ flies (\figpanel{fig:neural}{b}).  The corresponding
real-to-null-$95$th-percentile ratio has a median of $2.41$ across flies.  This
control establishes first-harmonic cue locking; it is not a null test of the
higher-order ratios from the half-wise phase-aligned estimator (S7).

The measured form factor, normalised to its first harmonic and taken across
flies, is $M_n/M_1 = 1,\,0.152,\,-0.016,\,-0.005,\dots$, with $41$ of
$46$ flies positive at $n=2$ and $M_2/M_1=0.152$ $[0.096,\,0.237]$.  It is
$0.148$ when all $115$ reference-aligned sessions from $60$ flies are retained.
Deconvolving the median $18.9^\circ$ split-half placement scale raises the
primary ratio to $0.171$ $[0.136,\,0.295]$; this is an upper-bound sensitivity,
because the split-half discrepancy also contains genuine compass drift (S7).
Two things follow.

First, this cue-referenced profile cannot determine the roll-off shape --- that
is, how the harmonic amplitudes $M_n$ decay with harmonic order $n$.  Only
the first two harmonics carry signal; on the single informative raw ratio, a
square bump of $W=162.5^\circ$ and a Gaussian of $\sigma=1.12$~rad fit exactly
equally (\figpanel{fig:neural}{c}).  Applying the upper-bound deconvolution
changes these to $160.3^\circ$ and $1.08$~rad without breaking the degeneracy
(\figpanel{fig:neural}{d}; S7).  In a
synthetic square-bump control, a placement error of about $20^\circ$ is enough
to favour the Gaussian; this is comparable to the measured split-half upper
bound of $18.9^\circ$ (interquartile range $10.1^\circ$--$33.6^\circ$).  This cue-referenced
dataset therefore cannot determine the roll-off shape.  The clean-ring
analysis above already rejects a sharp-edged square bump, whereas the contrast
manipulation in Sec.~\ref{sec:contrast} provides the stronger comparison among
candidate roll-off families.  The cue-referenced third harmonic is small,
negative and unresolved ($M_3/M_1=-0.016$).  A Gaussian bump would not produce
this sign, but residual phase-estimation or non-rigid alignment errors can
perturb a coefficient this small; the data do not identify its cause.  We
therefore do not interpret its sign as evidence about the intrinsic bump shape
(S7).

Second, the neural and behavioural second harmonics have the same order of
magnitude, but their point estimates do not satisfy the original smoothing
bridge.  The relation is
$\kappa_n/\kappa_1=(c_n/c_1)(M_n/M_1)$ (Eq.~\eqref{eq:factorization}), so
comparing the two ratios alone assumes a flat sensory stage.  The compass
gives $M_2/M_1=0.152$ $[0.096,\,0.237]$, or $0.171$ $[0.136,\,0.295]$ under
the upper-bound deconvolution; the fly
behaviour gives $\kappa_2/\kappa_1=0.36\pm0.05$ pooled over distance, or $0.40$
$[0.26,\,0.59]$ descriptively bracketed from the collapse slopes
(\figpanel{fig:neural}{d}).  The $0.40$ is the quotient of the unweighted point
slopes, $0.348/0.872$; the bracket uses endpoint quotients of their separate
whole-recording bootstrap intervals and is not a confidence interval for the
ratio.  The weighted point slopes of Sec.~\ref{sec:h2} give the larger
$0.375/0.666=0.56$.  The raw neural interval does not overlap the behavioural
bracket, whereas the upper-bound deconvolution does; neither comparison
establishes equality.  At the point estimates, even the deconvolved neural
ratio requires $c_2/c_1=0.40/0.171=2.34$.  Any non-negative sensory kernel
that is reflection-symmetric about the cue and nonincreasing with circular
distance satisfies $|c_2|\leq c_1$ (S7), so the point estimates lie outside
that class.  The uncertainty reaches its boundary, however, and the two
preparations do not provide a calibrated test of their ratio.  A non-unimodal
or sign-changing sensory kernel, or failure of another bridge assumption,
could explain the mismatch.
At the third harmonic both sides
are unresolved (\figpanel{fig:neural}{e}), and two quantities that are both
unresolved are not evidence for each other.  These are different animals in
different apparatus, so this is a compatibility check and not a within-animal
test: it constrains, but does not validate or directly refute, the proposed
bridge.

\subsubsection{Contrast sharpens the compass, and the sharpening predicts the
spectrum}
\label{sec:contrast}

The form factor has so far been a fixed quantity measured once.  In a subset of
the same archive it is manipulated: cue contrast is varied \emph{within fly}
across trials at three levels --- zero (darkness), low and high --- in $15$
sessions from $15$ flies.  That turns $M_n$ into a dependent variable.

The cue-referenced estimator cannot measure this effect, because its precision
depends on how visible the cue is --- exactly the variable being manipulated.
We therefore measure the bump in its own frame: regions are embedded on a ring
from their correlation matrix alone, the bump phase is read off per frame, and
the profile is accumulated relative to that phase, so the cue enters nowhere
(S7).  The width reported is $\sigma_{\mathrm{bump}}$,
fitted to the normalised harmonic coefficients under a Gaussian
parametrisation --- the shape the same data select below.

\begin{figure}[p]
\centering
\includegraphics[width=\textwidth]{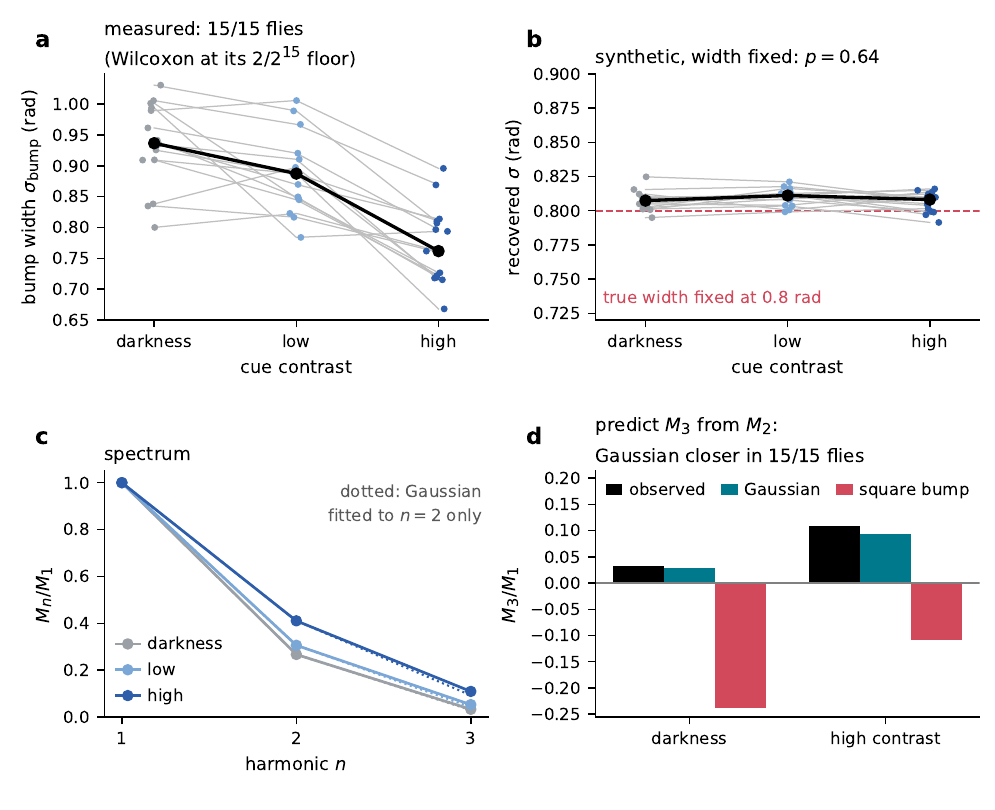}
\caption{
\textbf{Cue contrast sharpens the compass, and the sharpening predicts the
spectrum.}
\emph{Drosophila} compass imaging \citep{basnak2025multimodal,basnak2024dandi}.
\textbf{a}, Bump width in the cue-free bump frame, per fly, at three cue
contrasts.  Grey lines join the same animal and black is the median.  The bump
narrows as contrast rises, from a median of $0.94$~rad in darkness to
$0.76$~rad at high contrast.
\textbf{b}, The identical estimator applied to synthetic data whose width is
held \emph{fixed} at $0.8$~rad, rendered on each condition's own measured
phase trajectory with that condition's own noise.  When the truth does not
vary, no change in the estimate is detected ($p=0.64$), supporting the
interpretation of \textbf{a}.
\textbf{c}, The measured spectrum at each contrast.  Dotted curves are the
Gaussian implied by a width fitted to $n=2$ alone, extended to $n=3$.
\textbf{d}, The held-out-harmonic test, at the two extreme conditions.
Fitting the width to the second harmonic and predicting the third, the
Gaussian prediction is close to the observed value while the square bump predicts the wrong sign --- a
structural failure, not a matter of the fitted width.
Data: $15$ compass sessions, one per fly, each carrying all three contrasts. The
bump is measured in its own frame: ring positions come from
the region-by-region correlation matrix alone, so the cue never enters the
estimate.  The fly is the unit of replication and \textbf{a}, \textbf{b} and
\textbf{d} pair the three conditions within fly.  The shift in \textbf{a} runs
the same way in $15$ of $15$ flies, which puts its two-sided Wilcoxon $p$ at
the $2/2^{15}$ floor; \textbf{d} counts flies, not session-conditions, and the
Gaussian is closer in $15$ of $15$.  The synthetic control in \textbf{b} is
generated once per session-condition from that condition's own measured phase
path and residual noise.  A signal-to-noise control: degrading the
high-contrast recordings to the darkness noise level moves the width by
$0.0007$~rad, $0.4\%$ of the measured effect, and leaves all $15$ flies
narrower.  Full method in Note~S7.}
\label{fig:contrast}
\end{figure}

In the bump frame the width falls monotonically with contrast,
$\sigma = 0.937 \to 0.887 \to 0.762$~rad from darkness to high contrast, with
every one of the $15$ flies moving the same way (an exact Wilcoxon at its
$n=15$ floor of $6.1\times10^{-5}$; \figpanel{fig:contrast}{a}).  The bump's own amplitude does not follow
($0.541 \to 0.599$, $p=0.064$): contrast sharpens the compass profile without
a resolved increase in its amplitude.  Two controls support this interpretation.  A synthetic bump of
\emph{fixed} width, rendered on each condition's own measured phase trajectory
with that condition's own noise, returns a flat width ($p=0.64$;
\figpanel{fig:contrast}{b}).  And the narrowing is specific to sensory
reliability: stepping the closed-loop gain from $1$ to $2$ --- which destroys
the cue's validity as a heading reference while leaving its luminance and
geometry untouched --- yields no detected width change through the identical
pipeline; its detectable-effect calibration is given in S7.  Within these
manipulations, narrowing tracks cue contrast rather than closed-loop gain.

The width change predicts the harmonic changes, and the square bump fails by
sign.  Both higher harmonics rise with contrast: $M_2/M_1 = 0.267 \to 0.411$
and $M_3/M_1 = 0.033 \to 0.110$, each in $15/15$ flies with the exact
Wilcoxon again at its $n=15$ floor (\figpanel{fig:contrast}{c}).  Fitting each condition's width to the
\emph{second} harmonic alone and predicting the \emph{third}, a Gaussian
profile predicts $0.030$ against an observed $0.033$ in darkness and $0.093$
against $0.110$ at high contrast.  The square bump predicts $-0.238$ and
$-0.109$ --- the wrong sign, because $\sin(3W/2)$ has already turned negative
at the widths the second harmonic implies (\figpanel{fig:contrast}{d}).  Because the square bump fails structurally, the
informative comparison is against another smooth kernel: an exponential
roll-off with the same number of parameters and a heavier tail.  The Gaussian
is closer in $15$ of $15$ flies against the exponential and in $15$ of $15$
against the square bump, with the fly as the unit throughout.  With
every fly on the same side, the exact two-sided sign test is at its
$n=15$ floor of $6.1\times10^{-5}$: that number is fixed by the count of
$15$ and would be identical for an effect of any size, so the counts are
what the comparison rests on.  A shape-flexible check fits a single decay
exponent $\alpha$ per fly: in darkness the fitted exponent, $1.93$, is
indistinguishable from the Gaussian $\alpha=2$ against a calibrated null, and
it decreases to $1.72$ at high contrast --- a drift whose interpretation, via
the composition rule of Sec.~\ref{sec:general-kernel}, remains open (S7).

This measures the width of the compass bump, $\sigma_{\mathrm{bump}}$, under a
manipulation of cue contrast; it does not constrain the sensory kernel
$\sigma$, which is how precisely a bearing is localized.  For Gaussian kernels
the separation is impossible from behaviour alone: the behavioural spectrum
identifies only $\sqrt{\sigma^2+\sigma_{\mathrm{bump}}^2}$, so an independent
neural measurement of $\sigma_{\mathrm{bump}}$ is needed to separate the two
widths.  Here the behavioural and neural estimates come from
different preparations; their comparison does not establish a sensory width
or a calibrated upper bound (S7).  The gain from contrast is real but
modest: the third harmonic gains a factor of about three relative to the first
between darkness and high contrast.  A design seeking higher harmonics should
therefore use the brightest cues it can, but should not expect illumination
alone to deliver a third harmonic that the cue geometry does not.

\subsection{How far the measurement travels}

\subsubsection{A boundary response is strongly even-harmonic}
\label{sec:tetra}

The fly analysis uses a designed cue array.  A complementary test comes
from the arena boundary in the rummy-nose tetra recordings of
\citet{xue2023light}.  If a fish is at $(x_i,y_i)$, the bearing of the nearest
wall point --- the outward wall normal --- is
$\theta_{w,i}=\operatorname{atan2}(y_i,x_i)$.  Treating that direction as one
wall cue gives
\[
  F^{\mathrm{wall}}_{in}=e^{-\mathrm{i}n\theta_{w,i}},
  \qquad |F^{\mathrm{wall}}_{in}|=1
\]
at every harmonic.  The geometry therefore supplies flat leverage: the
arrangement itself neither favours nor annihilates any harmonic.  This does
not make the physical wall a point object; its angular extent and the animal's
filtering remain in the fitted response spectrum.  Wall and social terms were
estimated together.  In the nine near-wall strata containing neighbours, the
median magnitude of the wall second-harmonic coefficient was $2.17$, against
$0.39$ for the largest absolute social coefficient, a median per-cue ratio of
$5.1$.  More strictly,
the realised wall component --- including all of its harmonics --- exceeded
the summed social component in all nine near-wall strata (median ratio of
standard deviations $3.69$) and in $36$ of $38$ social strata overall (S6).
The wall is therefore the dominant fitted angular contribution near the
boundary, without implying that social effects are absent.

\begin{figure}[p]
\centering
\includegraphics[width=\textwidth]{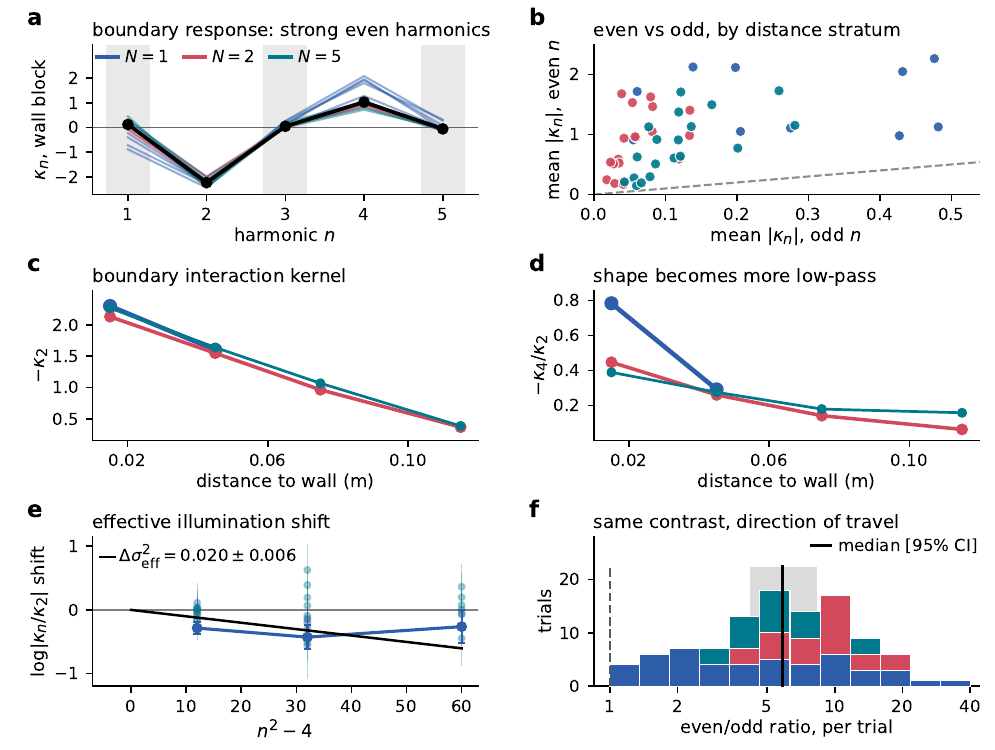}
\caption{
\textbf{The boundary response is strongly even-harmonic.}
Rummy-nose tetra \citep{xue2023light,xue2023data}.
\textbf{a}, Harmonic spectrum of the wall response, nearest stratum.  Faint
lines are individual conditions, coloured by group size, and the heavy line
the median.  Odd harmonics (shaded) are small and even harmonics large,
consistent with a strong axial wall-following component; the data do not
establish exact half-turn parity.
\textbf{b}, Mean $|\kappa_n|$ over even against odd harmonics, one point per
condition-by-distance stratum.  Every stratum lies well above the identity
line, but strata sharing a condition are not independent, which is why
\textbf{f} repeats the contrast at the trial.
\textbf{c}, The second harmonic falls with distance from the wall, tracing the
interaction kernel.  \textbf{d}, The normalized fourth harmonic falls too, so
the response becomes more low-pass with distance; the data do not identify
which sensory, cue or downstream factor causes that.
\textbf{e}, The illumination comparison, referenced to $n=2$ because the odd
harmonics are suppressed: the shift in $\log(|\kappa_n/\kappa_2|)$ between
each dim condition and its $50$~lx reference at $n=4,6,8$, against $n^2-4$,
with common slope $-\Delta\sigma_{\mathrm{eff}}^2/2$ under a Gaussian reading
of the composite spectrum.  The highlighted comparison --- solitary fish,
$0.5$ against $50$~lx, nearest stratum --- gives
$\Delta\sigma_{\mathrm{eff}}^2=0.020\pm0.006$, the line shown; but pooling all
twelve gives only $+0.004\pm0.002$ per decade of dimming, the sign unstable
across group size, so no sensory direction is claimed (S6).
\textbf{f}, The same even-to-odd contrast on the direction of travel, which
carries no head--tail ambiguity, with the trial as the unit: all $103$
eligible trials lie above a ratio of $1$ (dashed), median $5.8$, $95\%$
interval $[4.2,\,8.3]$ shaded.
Data: five illuminances, $0.5$ to $50$~lux, and
group sizes $N=1,2,5$, at $5$~Hz in a circular arena of radius $0.25$~m;
colour marks group size, keyed in \textbf{a}.  The cue is the wall, whose
nearest point lies outward along the radius: one cue, unit structure factor at
every harmonic.  \textbf{a}--\textbf{d} fit $n\leq5$ with social neighbours in
separate distance blocks, over the $48$ of $60$ condition-by-distance strata
holding at least $15\,000$ rows, from $15$ conditions
cut at $0$, $3$, $6$, $10$ and $26$~cm; $14$ of the $15$ clear it in the
nearest stratum drawn in \textbf{a}.  \textbf{e} refits to $n\leq8$ as a
wall-frame marginal, strata at $0$, $3$, $6$ and $12$~cm, $N=1$ and $N=5$
only.  Errors in \textbf{a}--\textbf{e} are cluster-robust on the trial file.
\textbf{f} uses the centred travel direction on rows above
$0.01$~m~s$^{-1}$, the $103$ of $106$ trial files holding at least $3000$
rows, and $10{,}000$ two-stage cluster bootstrap draws over the $15$
conditions.  Full method in Note~S6.}
\label{fig:tetra}
\end{figure}

The measured boundary spectrum is strongly even (\figpanel{fig:tetra}{a,b}).
Across the $14$ near-wall strata spanning solitary fish, pairs and groups of
five ($N=1,2,5$), the median coefficients through $n=5$ are
$(0.12,-2.24,0.05,1.03,-0.06)$: the second and fourth harmonics are large,
with $|t|$ reaching $43$ and $18$, while the first, third and fifth are small.
Extending to $n=8$ in the five solitary-fish strata, the even coefficients
alternate in sign and decay while the odd ones stay small throughout (S6).
Because the two tangential directions are geometrically equivalent, an
idealized axial wall-following component may satisfy
$R(\psi+\pi)=R(\psi)$, which removes all odd harmonics.  The no-penetration
constraint alone does not impose this symmetry because it distinguishes
inward from outward headings.  The observed even dominance therefore supports
a strong axial wall-following component, not an exact boundary-imposed parity
rule (S6).

Head--tail assignment errors suppress odd Fourier moments while leaving even
moments unchanged, so the posture-axis spectrum cannot by itself establish
parity.  Repeating the analysis with the unambiguous direction of travel
preserves strong even dominance: across $103$ trials, the median even-to-odd
ratio is $5.8$ $[4.2,\,8.3]$ (\figpanel{fig:tetra}{f}).  This control cannot establish exact parity,
however, because stationary confinement forces the speed-weighted mean radial
motion toward zero and can therefore suppress the first harmonic even without
a parity rule.  The data thus support strong even dominance, but not exact
annihilation of the odd harmonics (S6).

The supported conclusion is that the boundary response is strongly
even-dominated, and a first-harmonic description of wall avoidance
--- as in a simple repulsion rule --- does not capture its dominant structure.  The
even-harmonic results below do not rely on head--tail resolution: a half-turn
leaves every even heading feature unchanged, and the injection control finds
negligible fitted distortion (S6).

The distance dependence separates gain from shape as the theory says it should.
The second harmonic falls monotonically with distance from the wall,
$-\kappa_2=2.23,1.60,1.02,0.38$ across strata, tracing an interaction kernel
(\figpanel{fig:tetra}{c}).  The normalized fourth harmonic
$|\kappa_4/\kappa_2|$ also falls, from $0.54$ at the wall to $0.11$ furthest
from it (\figpanel{fig:tetra}{d}) --- the angular response becoming smoother
with distance from the boundary.

For illumination, the strong suppression of odd harmonics fixes $n=2$ as the
reference.  Extending the spectrum to $n=8$ then leaves $n=4,6,8$ to test,
rather than merely fit, the Gaussian relation
\[
 \log\!\left[
   \frac{|\kappa_n(L)|/|\kappa_2(L)|}
        {|\kappa_n(L_0)|/|\kappa_2(L_0)|}
 \right]
 =-\frac{n^2-4}{2}\,\Delta\sigma_{\mathrm{eff}}^2,
\]
for a dimmer level $L$ against the brightest level $L_0$ in the same
wall-distance stratum.  Here $\Delta\sigma_{\mathrm{eff}}^2$ is the
dim-minus-bright change
in the Gaussian-equivalent width of the complete even-harmonic spectrum.  It
equals a change in sensory variance only if illumination leaves cue extent
and all downstream filters unchanged.  The relation is not rejected, but the
shift is not robust: pooling gives
$\Delta\sigma_{\mathrm{eff}}^2=+0.004\pm0.002$ per decade of dimming, with the
sign unstable across group size (\figpanel{fig:tetra}{e}).  A
distance-interaction test intended to distinguish a perceiver-side sensory
change from illumination-dependent visibility of an extended boundary
resolves neither pattern (S6).  We therefore claim no direction for a sensory
effect of illumination.

\subsubsection{The boundary spectrum favours smooth decay over a pure sinc}
\label{sec:rolloff}

The wall fits estimate the observable composite coefficients $\kappa_n$, not
the sensory coefficients $c_n$ alone.  Although
$|F^{\mathrm{wall}}_{in}|=1$ gives equal geometric leverage at every harmonic,
the fitted sequence still combines cue extent, sensory filtering, and neural
or motor filtering.  We therefore compare candidate shapes for the composite
roll-off rather than treating a fitted width as a sensory measurement.

In each selected condition we fitted all four magnitudes $|\kappa_n|$ at
$n=2,4,6,8$ with marginal-standard-error weights; at least three had to exceed
two standard errors.  The three equally parametrised two-parameter forms were
\[
 G_n=Ae^{-n^2\sigma_{\mathrm{eff}}^2/2},\qquad
 S_n=A\left|\frac{\sin(n\Theta/2)}{n}\right|,\qquad
 E_n=Ae^{-n\gamma}.
\]
The sinc form is generated by a sharp-edged extended cue or by the square activity
bump of Harmonic Theory \citep{salahshourcouzin2026harmonic}; the exponential is a
smooth,
heavier-tailed comparator.  Head--tail flips leave even raw moments exactly
unchanged, and the injection control leaves the fitted even coefficients
essentially unchanged (S6).

Across the twenty conditions with at least three resolved even harmonics, the
Gaussian has a weighted residual sum at least two units below the sinc in
$17$, the sinc is lower by that margin in $2$, and one is tied.  The
heavier-tailed exponential is lower than the Gaussian in all twenty.  These
are descriptive comparisons rather than calibrated goodness-of-fit tests,
because the condition gate and cross-harmonic covariance are not propagated.
The pattern nevertheless disfavors a pure sharp-edged sinc over the resolved
orders.  The $n\leq8$ extension is a wall-frame marginal fit, so grouped-fish
coefficients can contain social angular structure; restricting the comparison
to solitary fish preserves the Gaussian--sinc and exponential--Gaussian
orderings (S6).

Within the Gaussian model, the median fitted scale is
$\sigma_{\mathrm{eff}}=0.445$~rad, with interquartile range
$[0.409,\,0.465]$.  This is a Gaussian-equivalent width of the composite
boundary spectrum, not a measurement of sensory width.  A Gaussian sensory
filter multiplied by a sinc cue or bump factor is mechanistically possible,
but that product has three free parameters $(A,\sigma,\Theta)$.  If restricted
to the three or four resolved coefficients, it would leave zero or one residual
degree of freedom; it was not fitted here, and these data cannot separate the
two widths reliably.  Doing so
requires an independent measurement or manipulation of cue extent or neural
filtering.

\subsubsection{Cross-preparation tests separate a portable measurement from
system-specific spectra}
\label{sec:replication}

The preceding analyses apply harmonic spectroscopy to target-directed
decisions in \emph{Drosophila} \citep{sridhar2021geometry}, compass activity
in \emph{Drosophila} \citep{kim2019generation}, and boundary responses in
rummy-nose tetra \citep{xue2023light}.  We next asked which parts of the
framework generalise across independent preparations, cue types and taxa.
Three public archives provide complementary tests.  Larval-zebrafish
phototaxis \citep{karpenko2020light} tests whether the occupancy estimator can
recover a cue-anchored landscape after separating it from laboratory
anisotropy.  Mouse head direction \citep{ajabi2023population} extends the
analysis to a mammal and provides a neural--behavioural consistency check,
although the reference-frame comparison is limited by the number of animals.
Two-beam phototaxis in a unicellular alga \citep{raikwar2025phototactic} tests
whether varying cue separation is sufficient by itself to produce a selection
transition.  These are tests of portability, consistency and specificity: the
measurement can travel across systems, but the spectrum and resulting
behaviour need not be universal.

Note~S8 gives the three analyses, controls and frame calibrations in full.
Every fit includes both sine and cosine quadratures at each harmonic because
an archive's angular zero need not coincide with the landscape axis.  This
matters directly in the larval-zebrafish data of
\citet{karpenko2020light}: the first-harmonic phase is $-85^\circ$, so a
cosine-only fit would miss almost the entire response.

\paragraph{Larval zebrafish phototaxis: cross-species validation of the
measurement.}
We first asked whether harmonic spectroscopy could recover a cue-anchored
behavioural landscape outside the fly dataset.  \citet{karpenko2020light}
recorded 18{,}322 orientation bouts from 47 larval zebrafish relative to a
virtual light source.  With one cue, $|F_n|=1$ in every observation, so the cue
arrangement gives equal geometric leverage to every harmonic
(Eq.~\eqref{eq:design-information}).  In the cue frame, the response is
dominated by a first harmonic, $|\kappa_1|=0.319$ (joint two-quadrature
Wald $\chi^2_2=28.2$, $p=7.4\times10^{-7}$), with no resolved higher harmonic
(all $p\geq0.19$).  The same fit in the laboratory frame gives a fourfold
weaker and unresolved first harmonic, $0.080$ ($\chi^2_2=1.05$, $p=0.59$).
A joint model confirms that these are distinct contributions: a cue-anchored
component of $0.40$ ($\chi^2_2=38.4$, $p=4.7\times10^{-9}$) remains after
accounting for a smaller room anisotropy of $0.19$ ($\chi^2_2=11.9$,
$p=0.0026$).
Thus, the dominant landscape follows the experimentally controlled cue rather
than a fixed laboratory direction (S8), independently validating the
occupancy-based measurement in another species.  Because this experiment has
only one cue, it does not test the compromise-to-selection transition; and
because no higher harmonics are resolved, it cannot distinguish alternative
decay profiles (S8).  If a Gaussian profile is assumed, the inferred scale is
$\sigma=0.88$~rad.  The fly signal-to-noise ratios give conditional
Gaussian-equivalent benchmarks of $1.06$ and $0.90$~rad (S1), not a fitted
width or uncertainty interval.  The rummy-nose-tetra boundary value of
$0.445$~rad \citep{xue2023light} is a composite scale for an extended wall
and is therefore not directly comparable to the point-like light-cue scales.

\paragraph{Mouse head direction: an informative limit of the frame control.}
The mouse data of \citet{ajabi2023population} allow the same estimator to be
applied to both measured head direction and a neural estimate of head
direction.  Across $719{,}633$ frames from 60 sessions, the cue-referenced
behavioural spectrum has $|\kappa_1|=0.437$ and $|\kappa_2|=0.143$.  However,
the laboratory-frame first harmonic is also large, $0.291$, or two thirds of
the cue-frame value.  Because cue positions are unevenly sampled and the 60
sessions come from only three mice, the data cannot reliably separate a
cue-anchored landscape from a fixed room preference.  Three-animal subsamples
of the zebrafish data show correspondingly wide variation in the same frame
comparison (S8).  The mouse result therefore neither confirms nor rejects
a cue-anchored landscape; instead, it shows that many frames cannot compensate
for too few independent animals when reference frames must be separated.  The
archive also includes a neural estimate of head direction generated by the
original authors' decoder.  Ajabi et al. trained this decoder to reconstruct
the animal's measured head direction from population activity.  The decoded
and measured spectra are similar in the same animals and reference frame,
$|\kappa_n|=(0.368,\,0.153,\,0.018)$ and
$(0.437,\,0.143,\,0.042)$ for $n=1,2,3$, respectively (S8).  This agreement
shows that our spectral analysis preserves the relationship encoded by the
authors' decoding pipeline, but, because measured head direction was the
decoder's training target, it is a consistency check rather than independent
neural evidence for the decision landscape.

\paragraph{A varied separation in a unicellular alga: a mechanistic contrast.}
The two-beam phototaxis data of \citet{raikwar2025phototactic} provide a
stringent contrast to the animal results: beam separation is varied across
five conditions spanning the same two-cue geometrical question, but
\emph{Chlamydomonas reinhardtii} is a single cell with no nervous system and
therefore has fundamentally different phototactic sensorimotor machinery.
Individual headings were measured for $5{,}300$--$13{,}100$ cells at beam
separations from $25^\circ$ to $135^\circ$.  The near-coincident and
intensity-imbalanced conditions determine the optical reference frame
independently of the result (S8).  In that frame, the second circular moment
about the bisector remains positive throughout, declining only from $+0.98$
to $+0.76$ $[0.67,0.82]$, with no sign change in $10^4$ file-level bootstrap
replicates.  A frame-independent calculation about the population's own mean
gives the same conclusion, never falling below $0.85$.  Across the five
released conditions, the observed distributions remain concentrated near the
bisector, with no resolved second lobe at either beam.  This is an informative mechanistic
contrast, not a failure of Harmonic Theory: Eq.~\eqref{eq:pair-sf} shows that
cue geometry can only filter harmonic landscape terms supplied by the
organism's response.  Here the released distributions remain single-lobed,
with no evidence in the measured second moment for the sign reversal associated
with the fly transition by $135^\circ$.  \citet{raikwar2025phototactic} report three
subpopulations at $162^\circ$, beyond the public conditions analysed here;
these data therefore show that separation alone does not force a fly-like
notch within the analysed range, not that the alga never transitions.

Together, these analyses separate portability of the measurement from
universality of the biological response.  The behavioural
compromise-to-selection transition is not fly-specific:
\citet{sridhar2021geometry} also reported it in larval zebrafish.  We could
not apply harmonic spectroscopy to that transition because the public release
available to us retains aggregate occupancy grids and position--target
trajectories but not direct body orientation or animal/trial identifiers.
Thus, the transition generalises beyond flies, whereas its harmonic-spectral
identification remains established only in the released fly recordings.

\section{Discussion}
\label{sec:discussion}

The central move of this paper is to stop treating the cue configuration as a
nuisance to be conditioned on and to start treating it as the instrument.
Write the arrangement of cues as a complex structure factor and the angular
decision landscape factors exactly as a diffraction amplitude does
\citep{alsnielsen2011elements}: symmetry acquires selection rules, and a design
gains a computable information content at every harmonic.  The consequences are
practical.  A configuration with $p$-fold symmetry is a harmonic comb filter,
passing only multiples of $p$; a swept pair separation is a tunable notch; and a
coefficient whose design leverage is zero should be reported as unmeasured, not
as absent.

The same reframing turns a familiar frustration into a quantitative prediction.
Compact summaries of interaction rules across taxa converge on first-harmonic
responses, and this has been read as a statement about animals.
Equation~\eqref{eq:detectability} says that much of this pattern is a statement about measurement: the sensory factor $e^{-n^2\sigma^2/2}$ suppresses the observable
amplitude as a Gaussian in $n^2$.  The fly signal-to-noise ratios illustrate
this loss of detectability, but do not estimate a width.  Under Gaussian sensory and neural filters only their combined width enters the spectrum, and reading a width back out of the ratios would depend also on the design's geometric leverage and on the sampling precision at each harmonic (S1).

\subsection*{What the neural data settle}

We showed that the neural half of the theory can be measured rather than
assumed.  The compass profile a cue evokes on the \emph{Drosophila}
protocerebral bridge has its own Fourier coefficients, and those coefficients
are what the behavioural spectrum is supposed to inherit.  Measured on a ring
recovered by split-half cross-validation and checked against a cue-shift
control, that profile carries a second harmonic of the same order as the
behavioural one, though the raw intervals meet only after a correction for
placement error (Sec.~\ref{sec:neural}).  Under the theory's own bridge
(Eq.~\eqref{eq:general-factorization}), even the corrected estimates would
require the sensory stage to pass the second harmonic more strongly than the
first, which no non-negative kernel that is reflection-symmetric about the cue
and nonincreasing with circular distance can do (S7).  The uncertainty reaches
its own boundary, so these data do not exclude that class at a calibrated level,
and a well-resolved ratio above one would still not identify its cause: a
centre--surround kernel, a non-unimodal profile, or a failed bridge assumption
could each produce it.  We report the agreement and the mismatch together;
combining them would obscure this discrepancy in the proposed bridge.

The compass contains a measurable third harmonic in its own frame.  It is unresolved in the
cue-referenced frame, and we draw no inference from two unresolved quantities
agreeing; in the bump's own frame it is resolved and it moves, rising with cue
contrast in $15$ of $15$ flies (Sec.~\ref{sec:contrast}).  What is missing is a
behavioural design with the leverage to read it.

The neural data also decide between the two profiles the theory allows.
Harmonic Theory gives its form factor for a square activity bump and for a
Gaussian one \citep{salahshourcouzin2026harmonic}; the square gives
$M_n\propto\sin(nW/2)/n$, a spectrum that oscillates in sign and decays only as
$1/n$.  The measured profile shows neither signature.  A static measurement in
the cue frame cannot settle the shape, because on one resolved ratio the two fit
identically; a manipulation can.  Under a within-fly contrast change, a width fitted to the \emph{second} harmonic predicts the \emph{third}: the square bump predicts the wrong sign, and the Gaussian is closer in $15$ of $15$ flies (Sec.~\ref{sec:contrast}).  The sharp-edged
description fares no better elsewhere: the sinc leaves a larger residual than
the Gaussian in $17$ of $20$ strata of the fish boundary response, a comparison
unaffected by head--tail sign errors because they leave every even
harmonic unchanged (S6), and on the ellipsoid-body ring the measured third
harmonic exceeds the square-bump prediction in thirty of thirty flies
(Sec.~\ref{sec:compassrule}).  The selection rule reads out there too: scenes
programmed with exact two- and four-fold symmetry suppress the forbidden moments
of the bump-to-scene offset in every animal, and elevation-broken twins at the
same azimuths restore them within fly.  These results span two species and
four measurement modalities.

The smooth form itself differs by preparation: the compass prefers a Gaussian
roll-off over an equally parametrised exponential in $15$ of $15$ flies, while
the fish boundary response prefers the heavier tail in all twenty conditions
(S6).  An extended wall is not a localized cue, so there is no reason its
effective kernel should match one.  The square bump is the analytically
convenient case and the theory gives the smooth one alongside it; the
measurements select the smooth kernel, and predictions that turn on oscillations
of $M_n$ with bump width do not survive that choice.

\subsection*{What discriminates a mechanism, and what does not}

A fitted coefficient is not by itself mechanistically diagnostic.  Of eight
generative rules passed through a common idealised closed-loop protocol and the
same form of occupancy analysis, three reproduce the observed second-harmonic
coefficient: winner-take-all, a leaky accumulator and a drift-diffusion process,
none of them parameterised as an explicit harmonic landscape.

The structure-factor collapse does better, because it tests a location the
geometry fixes in advance rather than an amplitude the fit is free to choose: it
excludes all six tested alternative parameterizations, and the held-out family
comparison extends the test to fitted families, excluding the null,
first-harmonic-only, winner-take-all and boxcar families when each is refitted
freely and scored on unseen animals (Sec.~\ref{sec:discrimination}).  Neither
test excludes everything.  Stochastic serial attention --- attend one target,
switch, repeat --- reproduces the collapse, and although the fixed rule's
$68.2^\circ$ notch lies below the primary $[81.9^\circ,113.2^\circ]$ interval, a
fitted serial family survives on turn persistence; graded softmax attention ties
the harmonic landscape in one cohort and outperforms it in the other.  Overall,
the data favour a graded angular response to cue geometry, the
harmonic account alone predicted the notch locations in advance, and a
purpose-built design could narrow the notch interval enough to distinguish the
remaining alternatives.

\subsection*{Limits of the present evidence}

Five limitations set the scope of these conclusions.

First, a pooled transition and an instantaneous landscape instability are not
the same quantity.  Passed through the observed centre--flank statistic, a
separately refitted first-harmonic model does produce a two-target crossing, but
too late, and none at all in the three-target cohort; the four-harmonic model
moves the prediction closer while the curve-wide error barely changes (S3.6).
The same caution applies to the boundary.  The spectrum establishes an
even-curvature reversal, but propagating the joint covariance gives intervals
broad enough to admit several crossings, and the three-target mean landscape has
appreciable slope at the geometric bisector.  The defensible claim is that
higher harmonics generate a reflection-even curvature reversal absent from a
first-harmonic landscape, not that these data fix a unique critical angle, and
the $n^2$ curvature weighting makes that a general caution for critical angles
read from fitted angular models (S3.5).

Second, the fitted higher harmonics do not transfer between designs.  The first
harmonic carries robustly across separations, geometries and disjoint sets of
animals; adding the fitted second gains nothing in one comparison and loses
accuracy in two others.  Part of this is diagnosable: with targets fixed in the
arena and the animal moving, the nominal middle target of the three-target
design is frequently not between the other two, so that configuration is not
mirror-symmetric in the animal's own frame.  The remaining causes are unresolved.

Third, illumination has no established directional effect on the spectrum.  The
pooled effect is small and its sign is unstable across group size, and the
distance-interaction test that would separate a perceiver-side sensory width
from a change in how much of an extended cue is visible resolves neither
pattern.  Only the composite shape is constrained.

Fourth, the occupancy channel (Eq.~\eqref{eq:occupancy}) requires the heading
distribution to equilibrate at the instantaneous cue configuration, in addition
to the gradient and constant-diffusion assumptions.  Our dynamical checks assess
this approximation for experimenter-placed cues (S4).  Extending it to cues
provided by other moving animals requires a separate test: in a self-propelled
group, members both respond to and generate one another's cue geometry.  Such
feedback can make conditional occupancy differ from the equilibrium response
to a fixed cue, as the moving-observer simulations already illustrate (S4).

Fifth, the comparative analyses separate portability of the measurement from
universality of any one spectrum.  The estimator recovers a cue-anchored first
harmonic in larval zebrafish; the mouse frame control remains unresolved; and
the unicellular alga shows no fly-like notch over the released range, the
reported splitting lying outside it \citep{raikwar2025phototactic}.  The
measurement generalises across systems, while the spectrum remains system-specific.  The behavioural
transition is not fly-specific --- it is reported in larval zebrafish as well
\citep{sridhar2021geometry} --- but its harmonic identification remains limited
to the released fly recordings.

\subsection*{The experiment this asks for}

For any prescribed arrangement and weighting of cues, the geometric leverage
at harmonic $n$, $\mathbb E|F_n|^2$, can be calculated before data
collection (Eq.~\eqref{eq:design-information}).  Designs can therefore be chosen
to emphasise the harmonics that separate competing hypotheses.  If animal
movement changes the apparent cue bearings, that expectation must instead be
averaged over the anticipated trajectories.  Three identical cues separated by
$120^\circ$ cancel the first and second cue-locked harmonics, so the response
begins at $n=3$; four identical cues separated by $90^\circ$ cancel harmonics
$n=1$ to $3$, so it begins at $n=4$.  Interleaving these symmetric arrays with a
densely sampled two-cue separation sweep in the same animals would isolate
higher harmonics by design, locate the notch precisely enough to test serial
attention, and determine whether a spectrum measured in one geometry predicts
responses to another without confounding differences between cohorts.

A related experiment has been run: the study supplying our fly cohorts also
reported compromise-to-selection bifurcations in larval zebrafish following
moving virtual conspecifics \citep{sridhar2021geometry}.  Those targets formed a
translating lateral array rather than an exact three-fold angular scene, and the
release lacks the body orientation and animal identifiers this estimator needs,
so a dedicated $p$-fold display is still required.

Two further measurements would reach the remaining latent factors.  Behaviour
identifies the product of cue-shape, sensory and neural filters, so at least one
factor must be varied or measured independently; virtual-reality preparations
that put the angular size of a virtual conspecific under experimental control
make this straightforward \citep{harpaz2024experience}.  Cue contrast, varied
within fly, already shows that one factor moves: it sharpens the compass bump in
$15$ of $15$ animals (Sec.~\ref{sec:contrast}), directly changing $M_n$.  It
leaves open whether the sensory coefficients $c_n(\sigma)$ stay fixed, since
they were not measured in the same manipulation.  A behavioural spectrum
recorded simultaneously would show how the product moves within animals, and an
independent cue-size or eccentricity manipulation would separate the two
factors.

\subsection*{A methodological point that generalizes}

For a one-dimensional overdamped gradient diffusion with a fixed cue
configuration, stationary occupancy is asymptotically as informative about the
drift as the full trajectory in the weak-response limit, so occupancy offers a
practical alternative to numerical differentiation.  Increment-based estimators
suffer an errors-in-variables bias of order $s^2/\delta t$, which grows as the
sampling interval decreases, whereas observation error in the occupancy
distribution enters through a known convolution that can be carried directly
into the likelihood.  In the fly recordings the two channels recover similar
normalised spectra (Sec.~\ref{sec:fdt}), which supports the relative
second-harmonic structure without establishing the exact
fluctuation--dissipation relation the ideal gradient model assumes (S1).  Any
field that infers interaction rules by regressing velocity changes on relative
position --- collective behaviour, chemotaxis, active matter --- has the second
channel available and is mostly not using it.

\subsection*{Conclusion}

Building on Harmonic Theory \citep{salahshourcouzin2026harmonic}, we
introduced harmonic spectroscopy of animal decisions: cue geometry becomes a
controlled spectral probe of behaviour and of neural representation.
Symmetric arrays fix which harmonics can appear, a swept pair drives a notch
through each of them, and the measured compass form factor selects the smooth
kernel over the square one.  This opens more questions than it settles: what
sets a species' spectrum; how its lines shift with state, learning and
development; how far into the nervous system the same selection rules reach.
All three are now open to experiment.  Cue geometry therefore offers a way to
resolve the decision spectrum of an orienting animal.

\section{Methods}
\label{sec:methods}

\subsection{Estimators}

\paragraph{Occupancy channel.}
For observation $i$ with heading $\theta_i$ and cue structure factors
$F_{in}$ (Eq.~\eqref{eq:structure-factor}) we fit the conditional circular
exponential family of Eq.~\eqref{eq:occupancy},
\begin{equation}
  \log p(\theta_i)=\sum_{n\in\mathcal N}\kappa_n
    \mathrm{Re}\!\left[F_{in}e^{\mathrm i n\theta_i}\right]
   +\sum_{m\in\mathcal M}\left[\alpha_m\cos m\theta_i+\beta_m\sin m\theta_i\right]
   -\log Z_i
  \label{eq:met-occupancy}
\end{equation}
by maximum likelihood.  The first sum is the cue spectrum; the second is a
laboratory-frame block that absorbs any orientation bias of apparatus, animal
or arena that does not follow the cues.  Omitting it is not conservative: in the
archived fly data \citep{sridhar2021geometry,godm2021data}, the lab-frame heading distribution
carries a second-harmonic resultant of $0.26$, which a fit without the block
reads as a cue response (S2).  With
several cue species present (social neighbours in distance strata, plus the
arena wall) each gets its own coefficient block and structure factor, so arena
geometry cannot leak into the social spectrum.  Quadrature grids,
optimiser settings and the constant-diffusion assumption behind
Eq.~\eqref{eq:met-occupancy} are given in S9.

\paragraph{Heading-error deconvolution.}
When heading is observed with wrapped-normal error of scale $s$, the
observed conditional density is the Gibbs density convolved with that law, and
we fit the convolved model directly.  Reported spectra are the uncorrected
fits; the deconvolved fits at $s\in\{0.05,0.15\}$~rad are a sensitivity
analysis (S2, S9).  This deconvolution removes heading noise from the
behavioural data; it is unrelated to the placement-error deconvolution
applied to the compass profile (S7).

\paragraph{Drift channel.}
Angular increments $\mathrm{wrap}(\theta_{k+1}-\theta_k)/\delta t$ are regressed
on the in-phase columns $-n\,\mathrm{Im}[F_ne^{\mathrm i n\theta_k}]$ of
Eq.~\eqref{eq:sde} and the quadrature columns
$-n\,\mathrm{Re}[F_ne^{\mathrm i n\theta_k}]$, plus the same laboratory
block and an intercept.  Rotational diffusion is estimated
from the residual variance.  Increments spanning a sampling gap are dropped.
In the fly recordings this residual-variance estimate requires an additional check:
one-frame heading changes are positively correlated, and it falls below the
diffusion measured from the mean-squared heading change by $32\%$ (two-target)
and $42\%$ (three-target).  The absolute two-channel comparison of
Eq.~\eqref{eq:fdt} is therefore reported at both values (S1).

\paragraph{Time-scales and replay.}
The second diffusion estimate is the short-lag slope of the mean squared
heading change; the relaxation time $\tau_1$ is an exponential fit to the
cue-frame heading decorrelation.  Both use windows declared before fitting
and are bootstrapped over whole recordings.  The open-loop replay plays back
each recorded bearing sequence with weight and frame time unchanged, and
simulates only the heading, from the cohort's fitted spectrum and measured
diffusion.  The replay isolates an open-loop approximation, but its comparison
with the idealised closed loop is not a controlled one-factor contrast.
Windows, fitting rules and the replay protocol are given in S4.

\paragraph{Uncertainty.}
Uncertainty is evaluated at the independent unit supplied by each archive,
never by treating frames as independent.  Fly behaviour is clustered or
resampled by recording; compass sessions are first aggregated within fly and
the primary intervals resample flies; fish-boundary and algal analyses use the
independent trial files or experiment files specified in S6 and S8; zebrafish
bouts are clustered by larva, and the mouse estimates are descriptive because
that archive supplies only three animals (S8).  The fly
separation bins and collapse regression are refitted inside each recording
bootstrap replicate.  Cluster-robust intervals use $t$ quantiles with $G-1$
degrees of freedom for $G$ independent units \citep{cameron2015practitioner}, and held-out comparisons are
paired across recordings.  Analysis-specific units, sandwich formulas, replicate
counts, seeds and interval conventions are collected in S9.

\subsection{Data}

We use the \emph{Drosophila} virtual-reality recordings of
\citet{sridhar2021geometry} and their public archive \citep{godm2021data}: 30
recordings with two virtual targets and 30 with
three, deterministically thinned to at most one observation per $50$-ms
acquisition-time bin while retaining the original timestamps, with the first
$0.5$\,s after each display switch
removed.  Heading is the measured closed-loop orientation, not a displacement
derivative.  Target bearings and distances are reconstructed from the
experiment and project databases in the archive.  The two-target cohort spans
$30$--$180^\circ$ of angular separation and the three-target cohort places a
third target on the bisector.  The two cohorts are disjoint sets of animals,
which makes the cross-geometry comparison a transfer rather than a refit.
The resulting table holds $639\,149$ frames.

Every binned analysis uses the same edges, $30^\circ$ to $180^\circ$ in
$15^\circ$ steps, with a minimum-occupancy rule that drops the lowest bin in
both cohorts.  Nine bins with midpoints $52.5^\circ$ to $172.5^\circ$ therefore
enter every collapse regression, and the bootstrap applies the same rule
inside each replicate (S9).

The boundary analyses use the rummy-nose tetra recordings of
\citet{xue2023light} and their public archive \citep{xue2023data} at five
illuminance levels and group sizes $N=1,2,5$,
with body orientation from the head--tail posture axis (S6, S9).  The
$N=25$ trials are excluded because the validated social design represents at
most four neighbours and would omit $20$ of $24$ neighbours for each focal
fish; S6 gives the scope criterion in full.  The
compass form-factor analyses use the public two-photon EPG compass-neuron
recordings of \citet{basnak2025multimodal,basnak2024dandi}, fixed at DANDIset
000289 version \texttt{0.240801.1040}.  The primary cue-referenced analysis
contains $101$ sessions from $46$ flies; ring reconstruction, session inclusion, the
cue-shift null and the placement-error deconvolution are specified in S7.  The
same archive supplies the two session sets read in the cue-free bump frame:
$15$ sessions from $15$ flies carrying a within-fly cue-contrast series, and
$14$ from $14$ flies carrying a closed-loop gain ladder (S7).
The scene-symmetry analyses use the public ellipsoid-body deposit of
\citet{kim2019generation}.  We reproduced the authors' preprocessing and
verified the resulting per-trial offsets against the released values (S7).

Three further public archives carry the tests of reach.
\citet{karpenko2020light} give bout-resolved orientation of larval zebrafish
relative to a virtual light source, with the relative bearing stored in the
archive rather than reconstructed: $18{,}322$ bouts from $47$ fish.
\citet{ajabi2023population} give mouse head direction with a visual cue on an
LED arena, together with a head direction decoded from the recorded
head-direction cell population; restricting to the epochs for which the archive
reports a cue angle leaves $719{,}633$ frames over $60$ sessions.
\citet{raikwar2025phototactic} give per-cell headings for the alga
\emph{Chlamydomonas reinhardtii} under two collimated LED beams, with beam
separation varied across five conditions from $25^\circ$ to $135^\circ$ and
$5{,}300$--$13{,}100$ cells per condition.  Frame calibrations, inclusion rules
and controls for all three are given in S8.

\subsection{Locating the second-harmonic notch}

The fitted second-harmonic coefficient $\lambda_2(\Delta)$ is known only at the
nine separation-bin midpoints, so a crossing is detected wherever the sign
changes between adjacent midpoints, and its location found by linear
interpolation.  When more than one crossing is present, the rule is the
\emph{first downward crossing}: upward crossings are never candidates, and the
downward crossing at the smallest separation is taken.  The rule contains no
target separation and cannot be tuned toward $90^\circ$ or $120^\circ$; it was
written after the point curves had been inspected, so it is design-derived
rather than blind.  A replicate with no downward crossing is
recorded as undefined and counted, never dropped.  Uncertainty is obtained
from the joint recording-cluster sandwich covariance of all nine bin fits;
$100{,}000$ independent Rademacher multipliers are applied to the recording
score vectors, the coefficients are reconstructed through the common bread
matrix \citep{cameron2008bootstrap,kline2012score}, and the same crossing rule is applied to every draw.  Because the curve is
sampled at $15^\circ$ spacing, decimal-valued point crossings are linear-interpolation
summaries rather than degree-scale precision; inference uses the full intervals.  The
rules we considered and rejected, and the crossing's stability under re-binning
and estimator family, are analysed in S3.  The design zero itself --- the
separation at which the measured
$\langle\mathrm{Re}\,F_2\rangle$ crosses zero (Eq.~\eqref{eq:structure-factor})
--- involves no behavioural data and is quoted as the realised prediction:
$90.3^\circ$ in the two-target cohort and $113.0^\circ$ in the three-target
cohort, the latter $7^\circ$ below the nominal $120^\circ$ of an idealised
triplet.

\subsection{Controls}

We evaluated five controls on the two fly cohorts.

\begin{enumerate}
\item \emph{Permutation null.}  Within each cohort and nominal cue layout, one
  of all $3!$ maps among the three recordings is sampled.  Each destination
  heading series is paired in retained row order with the mapped recording's
  realised target-bearing and target-weight series, truncating the pair to the
  shorter record.  This retains more than $99.85\%$ of rows while breaking the
  heading--geometry correspondence.  Fixed-point-free and unrestricted
  derangements are reported only as sensitivities (S3).
\item \emph{Laboratory anisotropy.}  Reported with and without the lab-frame
  block, plus a placebo rotation and an order-localisation analysis (S2).
\item \emph{Generative battery.}  Eight generative rules simulated through the
  same closed loop --- two virtual targets on a ring swept by the simulated
  animal's own translation, one turning rule per arm --- and passed through the
  same estimator.  Each rule is one parameter setting, not a whole model
  class; the model classes are tested separately by the held-out family
  comparison (S5).
\item \emph{Quasi-static approximation.}  The animal's own diffusion and
  relaxation time are measured directly.  A moving observer is then simulated
  at a range of speeds in an idealised two-target arena, with heading and
  position both generated, and each recorded cue sequence is replayed through
  the generative model, where the geometry is the animal's own and only the
  heading is generated (S4).
\item \emph{Stratification.}  Results are reported by cue distance and arena
  radius (S2).
\end{enumerate}

The remaining archives carry their own controls, stated with the analyses they
belong to: for the tetra recordings, a head--tail flip injection and a refit on
the unambiguous direction of travel (S6); for the two compass archives, a
within-trial time-shift null, split-half cross-fitting, a cue-shift null, a
placement-error deconvolution, and fixed-width, signal-to-noise-matched and
closed-loop-gain controls (S7); and for the three tests of reach, the
laboratory-frame and optical-frame calibrations (S8).

\subsection{Reproducibility}

Every analysis reported here is specified in the Supplementary Information.
Software versions, numerical settings and convergence criteria are given in
S9.

\section*{Acknowledgement}
The author thanks Iain D. Couzin for his generous support and insightful discussions.

We acknowledge AI assistance from Anthropic Claude and OpenAI ChatGPT Codex in
writing and editing, programming, mathematical and numerical analysis, figure
preparation, and reviewing the manuscript.  The author specified every
analysis, checked every derivation, verified every AI input, and takes full
responsibility for the content of the paper, including the parts prepared with
AI assistance.

\section*{Data and code availability}
No new animal data were collected.
Every analysis uses a published archive: the virtual-reality fly recordings of
\citet{sridhar2021geometry} \citep{godm2021data}; rummy-nose tetra recordings
\citep{xue2023light,xue2023data}; the two \emph{Drosophila} compass deposits
\citep{basnak2025multimodal,basnak2024dandi} and
\citep{kim2019generation,kim2019data}; larval zebrafish
\citep{karpenko2020light}; mouse head direction \citep{ajabi2023population};
and a unicellular alga \citep{raikwar2025phototactic}.  The analysis code will
be deposited in a DOI-bearing repository before publication.

\nocite{kubo1966fluctuation}
\nocite{liptser2001statistics}
\nocite{phipson2010permutation}
\nocite{ratcliff1978theory}
\nocite{risken1989fokker}
\nocite{usher2001time}
\bibliographystyle{unsrtnat}
\bibliography{references}

\end{document}